\documentclass[letterpaper,prb,preprint]{revtex4}
\usepackage{epsfig} \usepackage{subfigure} 
\usepackage{amsmath}
\usepackage{graphicx}
\graphicspath{ {./} }
\usepackage{bm}
\usepackage{subfigure}
\begin{document}
\title{A Systematic Study to Improve the Performance of SrCoO$_3$ as an Anion-Intercalation-Type Electrode for Supercapacitors Through Interface, Oxygen Vacancies, and Doping}
\author{Sadhana Lolla}
\affiliation{ National Graphene Research and Development Center, Springfield, VA, 22151, USA}
\author{Xuan Luo}
\affiliation{ National Graphene Research and Development Center, Springfield, VA, 22151, USA}
\date{\today}
\setlength{\parindent}{1cm}
\begin{abstract}
    Supercapacitors have recently gained popularity as possible energy storage systems due to their high cycling ability and increased power density. However, one of the major drawbacks of supercapacitors is that they have a low energy density, which makes them less effective than batteries. Herein, we explore different methods of increasing the supercapacitor performance of the perovskite SrCoO$_3$. We carry out first-principles calculations to systematically study how SrCoO$_3$$/$graphene interface, oxygen vacancies, and doping improve the performance of strontium cobaltite as an anion-intercalation-type supercapacitor. The results show that the SrCoO$_3$$/$graphene interface is relatively stable with a formation energy of 1.3 eV and is highly conductive, which makes it a promising material for supercapacitors. We also find that inducing oxygen vacancies in SrCoO$_3$ significantly increases the conductivity of this material. Results of doping calculations reveal that doping with Mo, V, P, and Nb all increase the stability and conductivity of SrCoO$_3$. We find that niobium is the most stable and most conductive of all four dopants. In addition, we find that vanadium is a very promising novel dopant for SrCoO$_3$ as an anion-intercalation-type supercapacitor electrode material.\par
\end{abstract}
\maketitle
\section{Introduction}
     Global warming is wreaking destruction on habitats around the world, polluting the air and water, and putting endangered species at risk. In light of this growing environmental crisis, renewable energy sources such as wind, solar, and hydrothermal energy are becoming crucial to the growth of sustainable clean energy\cite{saw2019}. In addition, over the past decade, concerns regarding environmental health have propelled the movement away from traditional gasoline-powered vehicles to more environmentally friendly such as electric vehicles (EVs) and hybrid electric vehicles (HEVs). However, both clean energy sources and EVs require fast and powerful energy storage devices \cite{wei2011}. Supercapacitors have the potential to fill this need due to their unique properties and widespread applications \cite{horn2019}.  \par
    Though the first patent for a supercapacitor was filed in 1957\cite{wang2012}, supercapacitors have only recently begun to attract widespread interest because of their fast charge-discharge rate, high power density, and long cycle life  \cite{tomar2018}. However, there are two main disadvantages facing the development of supercapacitors: they have high production costs and low energy densities (ED), which make them less effective than lithium ion batteries. The types of supercapacitors include traditional electric double layer capacitors (EDLCs) and pseudocapacitors. Of these two types, pseudocapacitors are generally preferred because they have a higher ED than EDLCs due to the reversible redox reactions between electrode materials and electrolytes  \cite{george2018}. However, the ED of pseudocapacitors still does not match the ED of batteries. Therefore, one popular method of improving supercapacitor performance has been to develop new electrode materials for pseudocapacitors. Materials used in pseudocapacitors include transition metal oxides (TMOs) \cite{yang2018controllable, yang2018, guo2018, xiong2013, nagarani2018, wang2014}, which have been extensively researched for this purpose. TMOs, specifically ruthenium oxide, are widely regarded as some of the most promising materials for pseudocapacitors\cite{wang2012}. However, ruthenium is an extremely expensive material and therefore electrodes made of ruthenium are not desirable. Recently, a third type of supercapacitor electrode has been proposed based on anion-intercalation mechanisms. These anion-intercalated electrode materials often produce higher energy densities than either pseudocapacitors or EDLCs because electrode reactions in the intercalation-type supercapacitors are similar to those in lithium-ion batteries\cite{liu2016}. \par
    Perovskite oxides of the form ABO$_3$ have the potential to be extremely effective anion-intercalation-type supercapacitor electrode materials due to their natural oxygen vacancies and defects\cite{ozolins2013}, which make them extremely conductive and give them the potential for both high energy density\cite{zhu2016} and high power density. Preliminary experimental studies have demonstrated high cycling ability and promising energy densities \cite{liu2018,lang2017,mo2018} of ABO$_3$ as a supercapacitor electrode. \par

Perovskite/graphene composites have the potential to further improve the energy density for supercapacitor applications, as graphene can enhance the conductivity of the perovskite oxide and the cluster structure of the perovskite can effectively reduce the agglomeration of graphene\cite{lang2017}. In addition, inducing oxygen vacancies in perovskite oxides can improve their performance as anion-intercalation-type supercapacitor electrodes. Furthermore, doping the perovskite oxide at the B-site increases the stability and conductivity of the complex, especially of cubic perovskite oxides \cite{aguadero2012, zhu2016, mo2018}. However, we are currently unaware of studies that have focused on perovskite oxides combined with graphene for supercapacitor applications. Therefore, our objective is to determine whether the performance of perovskite oxides can be improved through various methods in order to increase the energy density and/or power density of supercapacitors.  \par

We choose SrCoO$_3$ as our perovskite oxide because of its high Oxygen Evolution Rate (OER)  \cite{jia2017} and its behavior as an extremely active catalyst\cite{calle2015}, which means that oxygen vacancies can be created very easily in this material. Using first-principles calculations, we explore the conductivity and stability of an SrCoO$_3$$/$graphene interface. We also induce oxygen vacancies in SrCoO$_3$ to observe their effect on the conductivity of the perovskite for supercapacitor applications. In addition, we conduct doping calculations with Mo, V, P, and Nb to determine which dopant is the most effective in terms of conductivity and stability. We calculate band structure, DOS, and formation energy for all of the compounds.  \par
The rest of this paper is structured as follows: first, we will discuss the computational methods and material selection process and we will then discuss the results of our DFT calculations and their impact on the field of supercapacitors as a whole.  \par

\section{Methods}
\subsection{Computational Methods}
All first-principle density functional theory (DFT) calculations were conducted using ABINIT \cite{gonze2009}. The Generalized Gradient Approximation (GGA) and the Perdew-Burke-Ernzerhof (PBE) \cite{perdew1996} exchange-correlation functionals were utilized to determine the electronic structure properties of various materials. The pseudopotentials were based on the Projector Augmented Wave (PAW)\cite{blochl1994} method and the projectors were generated using AtomPAW\cite{holzwarth2001}.
\begin{center}
\begin{table}
\caption{Electronic Configuration and PAW radius cutoff used to generate PAW
pseudopotentials}
\begin{tabular}{ccc}
\hline
\hline
Element & Electronic Configuration (core/val) & PAW radius cutoff (a.u.) \\
\hline
Strontium & [Ar ${3d}^{10}$] ${4s}^2$ ${4p}^6$ ${5s}^1$ ${4d}^{1}$ & 2.2 \\
Cobalt & [Ne] ${3s}^2$ ${3p}^6$ ${4s}^1$ ${3d}^8$ & 2.1 \\
Oxygen & [He] ${2s}^2$ ${2p}^4$ & 1.4 \\
Carbon & [He] ${2s}^2$ ${2p}^2$ & 1.5 \\
Molybdenum & [Ar ${3d}^{10}$] ${4s}^2$ ${4p}^6$ ${5s}^1$ ${4d}^{5}$ & 2.2 \\
Phosphorus & [Ne] ${3s}^2$ ${3p}^3$ & 1.9 \\
Niobium & [Ar] ${3d}^{10} $ ${4s}^2$ ${4p}^6$ ${5s}^1$ ${4d}^4$ & 1.4 \\
\hline
\hline
\end{tabular}
\end{table}
\end{center}
\subsection{Convergence}
The kinetic energy cutoff and Monkhorst-Pack k-point grids were converged for all materials. The self-consistent field (SCF) total energy tolerance was set as 1.0 $\times$ $10^{-10}$ Ha. Once this tolerance was reached twice consecutively, the SCF iterations were terminated. The kinetic energy cutoff and Monkhorst-Pack k-point grids were considered converged when the differences in total energies were less than 1.0 $\times$ $10^{-4}$ Ha (about 3 meV) twice consecutively. The converged values for all materials can be found in Table \ref{tab:ecutkpt}. \par
\begin{center}
    \begin{table}
    \caption{Converged values of the kinetic energy cutoff and Monkhorst-Pack k-point grids for each material.}\label{tab:ecutkpt}
\begin{tabular}{ccc}
\hline
\hline
Material & Energy Cutoff (Ha) & k-mesh\\
\hline
SrCoO$_3$ & 20 & 4 $\times$ 4 $\times$ 4 \\
Graphene & 19 & 4 $\times$ 4 $\times$ 1 \\
Mo & 13 & 6 $\times$ 6 $\times$ 6 \\
P & 23 & 8 $\times$ 8 $\times$ 8 \\
Nb & 17 & 6 $\times$ 6 $\times$ 6 \\
V & 13 & 6 $\times$ 6 $\times$ 6\\
\hline
\hline
\end{tabular}
    \end{table}
\end{center}
\subsection{Relaxation}
The converged values for the kinetic energy cutoff and k-mesh were used to fully relax all systems studied. All structures were relaxed using the Broyden$-$Fletcher$-$Goldfarb$-$Shanno algorithm. The SCF iterations were terminated when the Hellman-Feynman forces were less than 5.0 $\times$ ${10}^{-6}$ Ha/Bohr twice consecutively. Structures were considered fully relaxed when the maximal force was less than 5 $\times$ ${10}^{-5}$ Ha/Bohr. \par
\subsection{Materials}
When used in supercapacitor applications, perovskite oxides usually follow the structure ABO$_3$. We choose SrCoO$_3$ as our perovskite because of its high OER\cite{jia2017}. Other theoretical studies have also predicted that strontium cobaltite has the highest OER ability among all perovskites through DFT calculations \cite{calle2015}. This makes SrCoO$_3$ a good candidate for supercapacitors because oxygen vacancies increase conductivity, especially in anion-intercalation supercapacitors \cite{alexander2019}. \par
The cubic structure of the perovskite SrCoO$_3$ was chosen as the base material for all calculations due to its high conductivity.

\subsubsection{Interface}
Experimental studies have speculated that graphene has the potential to increase the stability conductivity of perovskite oxides \cite{lang2017}. However, we are unaware of any experimental or theoretical studies to date that have focused on the effects of graphene on perovskite oxides for supercapacitor applications. In this study, we simulate an interface by combining an SrCoO$_3$ supercell with graphene. To do this, the kinetic energy cutoff was calculated for both pristine SrCoO$_3$ and pure graphene and the larger of the two was used for interface calculations. These values can be found in Table \ref{tab:ecutkpt}. The relaxed lattice constant for SrCoO$_3$ was chosen for the interface in order to best analyze the effects of graphene on strontium cobaltite. To obtain the layer-by-layer distance $d$ between SrCoO$_3$ and graphene, we systematically increase $d$ in increments of 0.5 Bohr and compare the total energy. The layer-by-layer distance with the lowest total energy is used in band structure and PDOS calculations. \par

\subsubsection{Oxygen Vacancy}
Numerous studies have demonstrated that oxygen vacancies improve the conductivity of compounds in supercapacitor applications \cite{yang2018, zhai2014, xiang2017}. This is especially true for perovskites of the form ABO$_3$ where B is a transition metal because oxygen vacancies change the geometric and chemical properties \cite{yang2018} of TMOs. To the best of our knowledge, oxygen vacancies have not been theoretically studied in perovskite oxides for supercapacitor applications. In the present study, we induce oxygen vacancies in strontium cobaltite to determine the mechanism by which vacancies improve the conductivity of this perovskite oxide. We simulated one oxygen vacancy in a 2 $\times$ 2 $\times$ 1 supercell by removing an oxygen. This structure was then relaxed and the band structure, PDOS, and formation energy were calculated. \par
Two oxygen vacancies were created by taking out an oxygen and its nearest neighbor. The structure was fully relaxed again and the above procedures were repeated for the complex with two vacancies. \par

\subsubsection{Doping}
We choose four different dopants for SrCoO$_3$: Mo, V, P, and Nb because they have been shown to increase the OER of SrCoO$_3$ and other similar perovskites. Substitutional doping of 25\% was achieved at the B-site by replacing one cobalt with a dopant in a 2 $\times$ 2 $\times$ 1 supercell. The kinetic energy cutoff was calculated for each dopant and for pristine SrCoO$_3$ and the maximum of the two was chosen for the doping calculations. These values are displayed in Table \ref{tab:ecutkpt}. Previous experimental studies have explored Mo and Nb as dopants for SrCoO$_3$ as an anion-intercalation-type supercapacitor and concluded that both Mo and Nb increase the conductivity of strontium cobaltite \cite{tomar2018,li2017niobium}. We use Mo and Nb as dopants in this study to examine how they improve the performance of SrCoO$_3$ as an electrode material for anion-intercalation-type supercapacitors. P-doped SrCoO$_3$ has previously been studied as a water oxidation electrocatalyst and has been shown to increase the OER of SrCoO$_3$\cite{zhu2016}. We use phosphorus as a dopant to determine whether phosphorus can be as effective as the other metallic dopants for SrCoO$_3$. Vanadium-doped transition metal oxides have been explored as possible electrode materials\cite{yang2013} for pseudocapacitors. In this study, we analyze the conductivity and stability of the V-doped SrCoO$_3$ to determine if it is also an effective dopant for perovskite oxides for anion-intercalation-type supercapacitors. \par

\subsection{Band Structure}
\begin{figure}
\subfigure[]{\includegraphics[width=4cm]{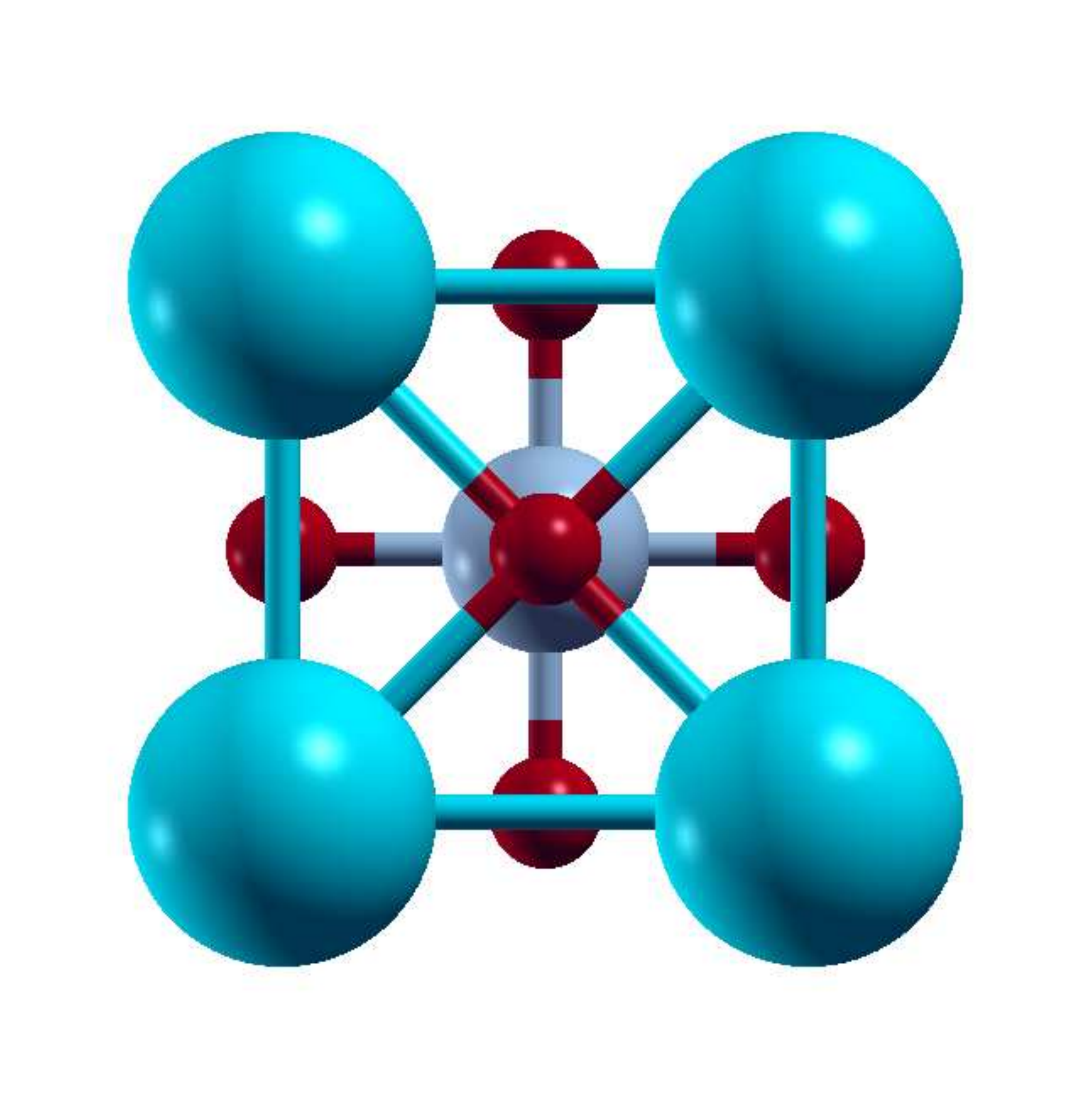}}
\subfigure[]{\includegraphics[width=7cm]{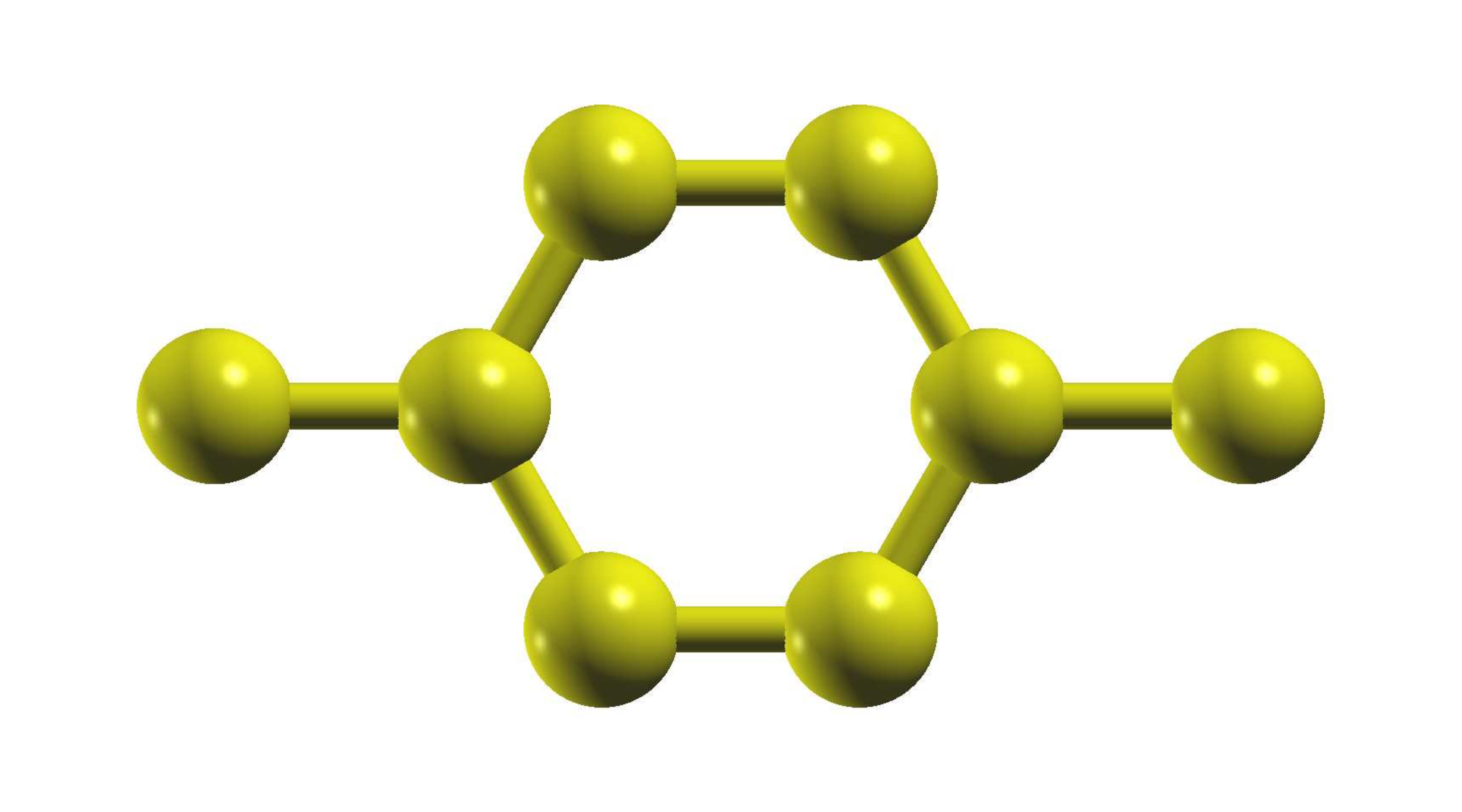}}
\subfigure[]{\includegraphics[width=6cm]{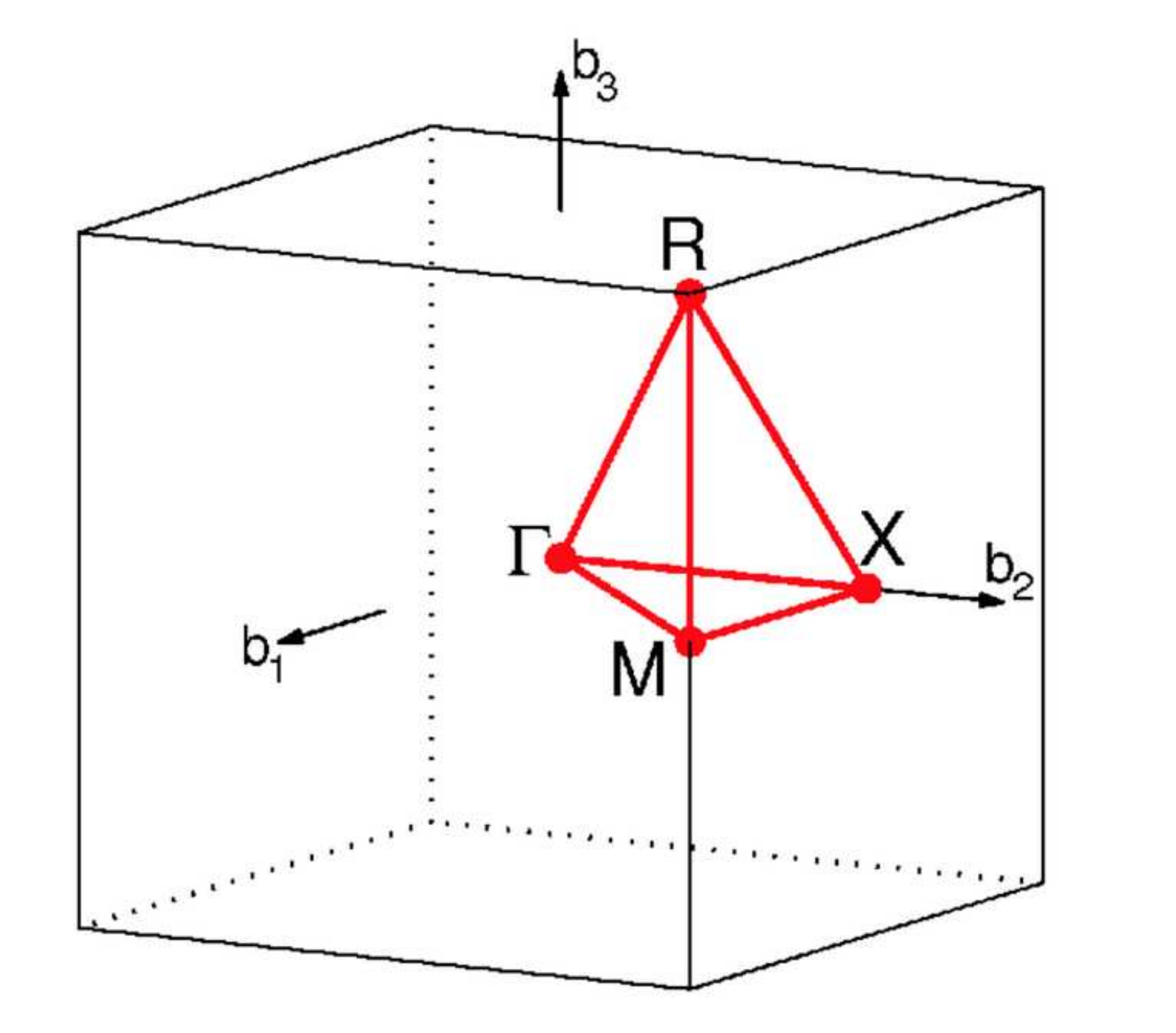}}
\subfigure[]{\includegraphics[width=8cm]{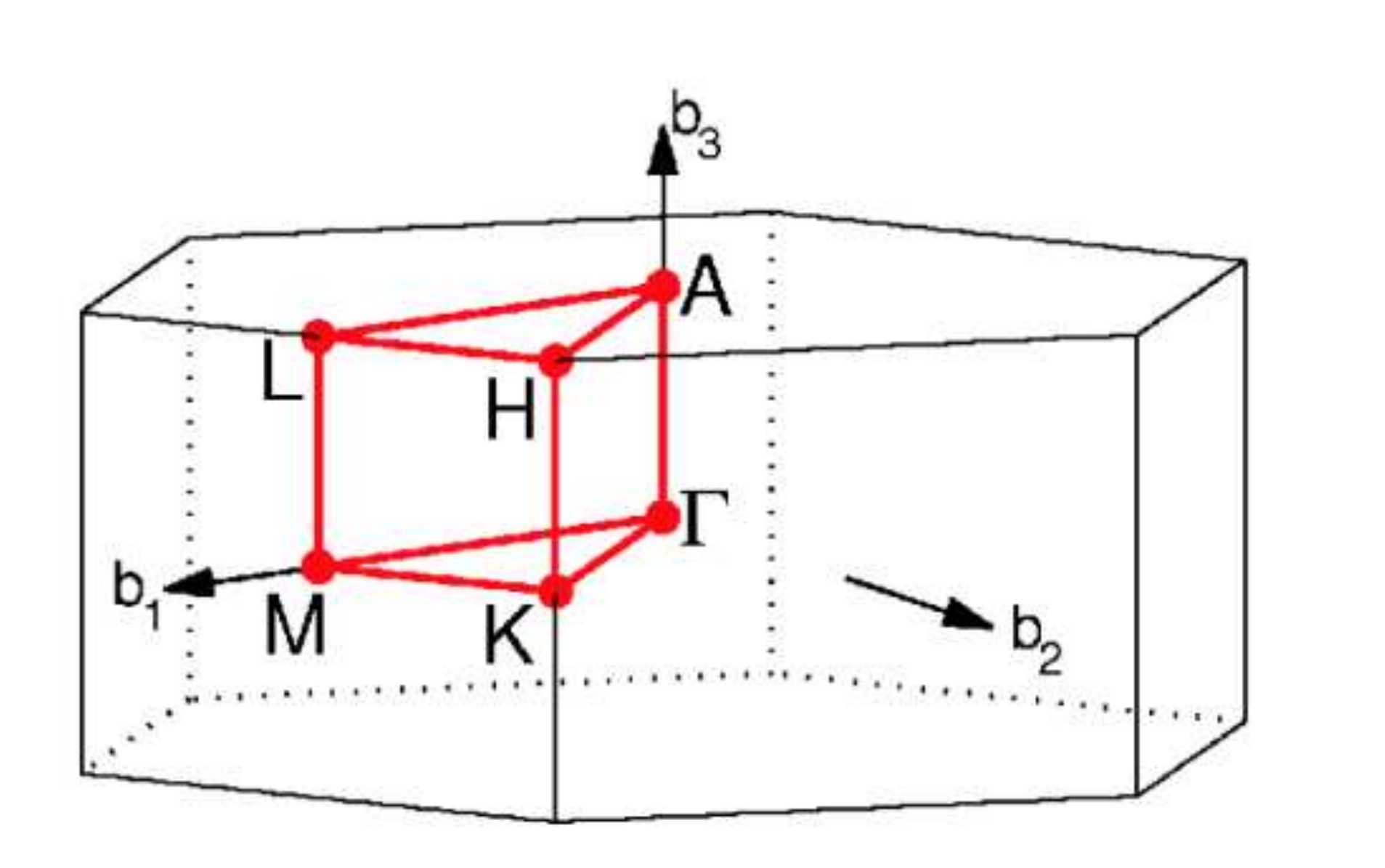}}

\caption{(a) Aerial view of SrCoO$_3$ simple cubic crystal structure. (b) Aerial view of hexagonal monolayer graphene. (c) Simple cubic high symmetry k-points\cite{setyawan2010}. (d) Hexagonal high symmetry-kpoints in the first Brillouin zone\cite{setyawan2010}.}
\label{bandstructure}
\end{figure}

Using the cubic atomic structure of SrCoO$_3$ shown in Figure \ref{bandstructure} (a), the band structure was calculated for the interface, vacancy, and doping calculations. To transform the lattice vectors $a_1$, $a_2$, and $a_3$ into the reciprocal space lattice vectors $b_1$, $b_2$, and $b_3$, Eqs. (1)-(3) were used. High symmetry k-points were then selected to comprehensively sample the first Brillouin zone. For doping and vacancy calculations, the high symmetry k-points for simple cubic structures were chosen and are: $\Gamma$ (0, 0, 0), X(0, 0.5, 0), M(0.5, 0.5, 0.0), and R(0.5, 0.5, 0.5). The first Brillouin zone and high symmetry k-points for simple cubic structures are shown in Figure \ref{bandstructure} (b). When plotting the band structures, each band contained two electrons and four additional conduction bands were used. \par
Likewise, the band structure was also calculated for pure hexagonal monolayer graphene, which is shown in Figure \ref{bandstructure}(c). Eqs. (1)-(3) were used to transform the lattice vectors into the reciprocal space lattice vectors, and the high symmetry k-points for hexagonal structures were used: $\Gamma$ (0, 0, 0), K(1/3, 2/3, 0), and M(0, 0, 0). Figure \ref{bandstructure}(d) shows the first Brillouin zone and hexagonal high symmetry k-points. \par
We plotted the fat-band structures for the doped complex structures to analyze the contributions of dopants to the overall band structure of the compound. Similarly, we also calculate the fat-band structures for SrCoO$_3$ with oxygen vacancies to study the contributions of the Oxygen $2p$ orbital to the band structure of the complex. \par

\begin{equation}
    \vec{b_1} = 2\pi\frac{\vec{a_2} \times \vec{a_3}}{\vec{a_1} \cdot (\vec{a_2} \times \vec{a_3})}
    \label{eqn:recspace1}
\end{equation}
\begin{equation}
        \vec{b_2} = 2\pi\frac{\vec{a_3} \times \vec{a_1}}{\vec{a_2} \cdot (\vec{a_3} \times \vec{a_1})}
        \label{eqn:recspace2}
\end{equation}
\begin{equation}
        \vec{b_3} = 2\pi\frac{\vec{a_1} \times \vec{a_2}}{\vec{a_3} \cdot (\vec{a_1} \times \vec{a_2})}
        \label{eqn:recspace3}
\end{equation}

\subsection{Density Of States}
Total DOS (TDOS) and projected DOS (PDOS) were calculated for the SrCoO$_3$$/$graphene interface, oxygen vacancy, and doping complex systems. The TDOS was used to analyze the overall conductivity of the complex and the PDOS was analyzed to find the contributions of each orbital to the conduction process. We plot the PDOS of the $3d$ orbital for Sr, the $3d$ orbital for Co, the $2p$ orbital for Oxygen, and the $d$ orbitals for all the dopants except Phosphorus, which lacks a $d$ orbital, so the $p$ orbital was used instead. \par

\subsection{Charge Transfer/Density}
We calculate the charge density for structures with oxygen vacancies in order to determine where the charge accumulates after the vacancy is induced. To do this, the converged charge density of the complex structure was obtained by calculating the total energy using the converged values for the kinetic energy cutoff and Monkhorst-Pack k-point grids along with the relaxed lattice constants. This total energy calculation is then used to generate isosurface diagrams for both one vacancy and two vacancy calculations. \par
We also calculate the charge transfer for the SrCoO$_3$$/$graphene interface using the formalism in Eq. (\ref{eqn:chargetr}). To calculate the charge transfer, the converged charge density was obtained through the total energy calculation. The total energy is calculated using the converged kinetic energy cutoff, Monkhorst-Pack k-point grids, and layer-by-layer distance $d$. The charge density of pristine SrCoO$_3$ and pure graphene are subtracted from the charge density of the interface to determine the electrons gained or lost. The red regions in the isosurface diagrams represent the charge accumulation and the blue regions represent the charge depletion.  \par
\begin{equation} Q_{transfer} = Q_{interface} - Q_{SrCoO3} - Q_{graphene}
    \label{eqn:chargetr}
\end{equation}

\subsection{Formation energy}
The formation energies for doping, oxygen vacancy, and interface complex systems were calculated. A lower formation energy indicates a more stable compound, and a negative formation energy is preferred because it indicates that an exothermic reaction has occurred. The formation energy for all complex systems were calculated using Eq. (\ref{eqn:formation energy}), where $E_f$ represents the formation energy. \par
\begin{equation} E_{f} = E_{complex} - E_{A} - E_{B}
    \label{eqn:formation energy}
\end{equation}
For the SrCoO$_3$$/$graphene interface, the total energy of pure SrCoO$_3$ and pure graphene were calculated using the converged values for the kinetic energy cutoff and k-point grids. In this case, $E_{complex}$ represents the total energy of the interface. $E_A$ represents the total energy of 1 $\times$ 1 $\times$ 2 SrCoO$_3$, and $E_B$ represents the total energy of 2 $\times$ 2 graphene. The formalism is displayed in Eq. (\ref{eqn:interface_energy}).\par

\begin{equation} E_{f} = E_{complex} - 2 \times E_{SrCoO3} - 4 \times E_{graphene}
    \label{eqn:interface_energy}
\end{equation}

In order to determine the formation energy of strontium cobaltite with vacancies, the total energy of oxygen gas was first calculated. $E_{complex}$ represents the formation energy of the complex with either one or two $V_O$. $E_A$ is the total energy of 2 $\times$ 2 $\times$ 1 SrCoO$_3$. $E_B$ is equal to $n$ multiplied by the total energy of one oxygen atom, where $n$ is the number of vacancies. The equation for the defect formation energy of the vacancy calculations can be found in Eq. (\ref{eqn:vacancy_energy}).\par

\begin{equation} E_{f} = E_{complex} - 4 \times E_{SrCoO_3} + n \times E_{O_2}/2
    \label{eqn:vacancy_energy}
\end{equation}

To obtain the formation energy of doped SrCoO$_3$, the total energies of the dopants and cobalt were calculated, along with the total energy of pure SrCoO$_3$. $E_{complex}$ represents the total energy of the doped compound, which was calculated using the converged kinetic energy cutoff, Monkhorst-Pack k-point grids, and relaxed lattice constants. As the dopant is being substituted for a cobalt atom, we add the total energy of the cobalt atom and subtract the total energy of the dopant atom. Eq. (\ref{eqn:dopant_energy}) shows the formalism for the formation energy of the doping calculations.

\begin{equation} E_{f} = E_{complex} - 4 \times E_{SrCoO_3} + E_{Co} - E_{dopant}
    \label{eqn:dopant_energy}
\end{equation}

\section{Results and Discussion}
In this section, we use DFT calculations to analyze the electronic and structural properties of an SrCoO$_3$$/$graphene interface, oxygen vacancies, and doping calculations as they apply to supercapacitor performance. \par
\subsection{Interface}
Previous studies have suggested that graphene has the potential to improve supercapacitor performance. Yang et al\cite{yang2018}, for example, has demonstrated that creating a Co$_3$O$_4$$/$graphene interface increased the energy density dramatically from 32 W h$/$kg to 43.1 W h$/$kg. Lang et al.\cite{lang2017} also suggested that a perovskite$/$graphene interface has the potential to further increase the conductivity and stability of this material. Therefore, our goal is to theoretically study a perovskite/graphene interface to understand why graphene improves perovskite performance for supercapacitors. \par

An SrCoO$_3$$/$graphene interface was formed by 1 $\times$ 1 $\times$ 2 SrCoO$_3$ and 2 $\times$ 2 graphene. The lattice parameters of the relaxed SrCoO$_3$ were used for the interface calculations. In lieu of relaxation, the optimal layer distance $d$ between the perovskite and graphene was determined by incrementally increasing $d$ by 0.5 Bohr and calculating the total energy. The layer distance with the lowest total energy was 5.5 Bohr, so this structure was used for DOS and formation energy calculations. We conclude that this value is reasonable because the graphene layer distance is around 6.6 Bohr\cite{affoune2001}. \par

The lattice constants of the relaxed compounds are displayed in table \ref{interface_structure}. The percentage error between the relaxed lattice constants and the published theoretical and experimental lattice constants for SrCoO$_3$ is always less than 1\%. Furthermore, the relaxed lattice constants for graphene calculations fall within the range indicated by Girit et al\cite{girit2009graphene}, indicating that our relaxation calculations are reliable for both SrCoO$_3$ and graphene. \par

\begin{center}
\begin{table}
    \caption{Relaxed Lattice Constants of SrCoO$_3$ and graphene} \label{interface_structure}
    \begin{tabular}{ccccc}
        \hline
    \hline
        Compounds & Calculation Method & a (Bohr) & b (Bohr) & c (Bohr) \\
        \hline
        SrCoO$_3$ & GGA (this study) & 7.21 & 7.21 & 7.21 \\
                & Experiment \cite{wang2012porous} & 7.28 & 7.28 & 7.28 \\
                & GGA \cite{jia2017} & 7.19 & 7.19 & 7.19 \\
        Graphene & GGA(this study) & 4.62 & 4.62 & \\
             & Experiment \cite{girit2009graphene} & 4.35 - 4.72 & 4.35 - 4.72 &   \\
                 & GGA \cite{giovannetti2008} & 4.62 & 4.62 & \\
        \hline
        \hline  
    \end{tabular}
\end{table}
\end{center}

To calculate the formation energy of the SrCoO$_3$$/$graphene interface, the total energy of pure SrCoO$_3$ and pure graphene were first calculated using the relaxed lattice constants given in Table \ref{interface_structure} and the converged kinetic energy cutoff and Monkhorst-Pack k-point grids given in Table \ref{tab:ecutkpt}. The formalism displayed in Eq. (\ref{eqn:interface_energy}) was then applied to calculate the formation energy of the interface, which is 1.3 eV. Previous theoretical studies such as Hussain et al\cite{hussain2019} have studied interactions between transition metal oxides (TMOs) and graphene for supercapacitor applications and determined the formation energy of CuO$/$graphene as 0.3 eV and FeO$/$graphene as 3.7 eV. Since the formation energy of the SrCoO$_3$$/$graphene interface falls between these two values, it can be concluded that the formation energy for this interface is reasonable and that the interface is relatively stable. \par

\begin{figure}
\subfigure[]{\includegraphics[width=5cm]{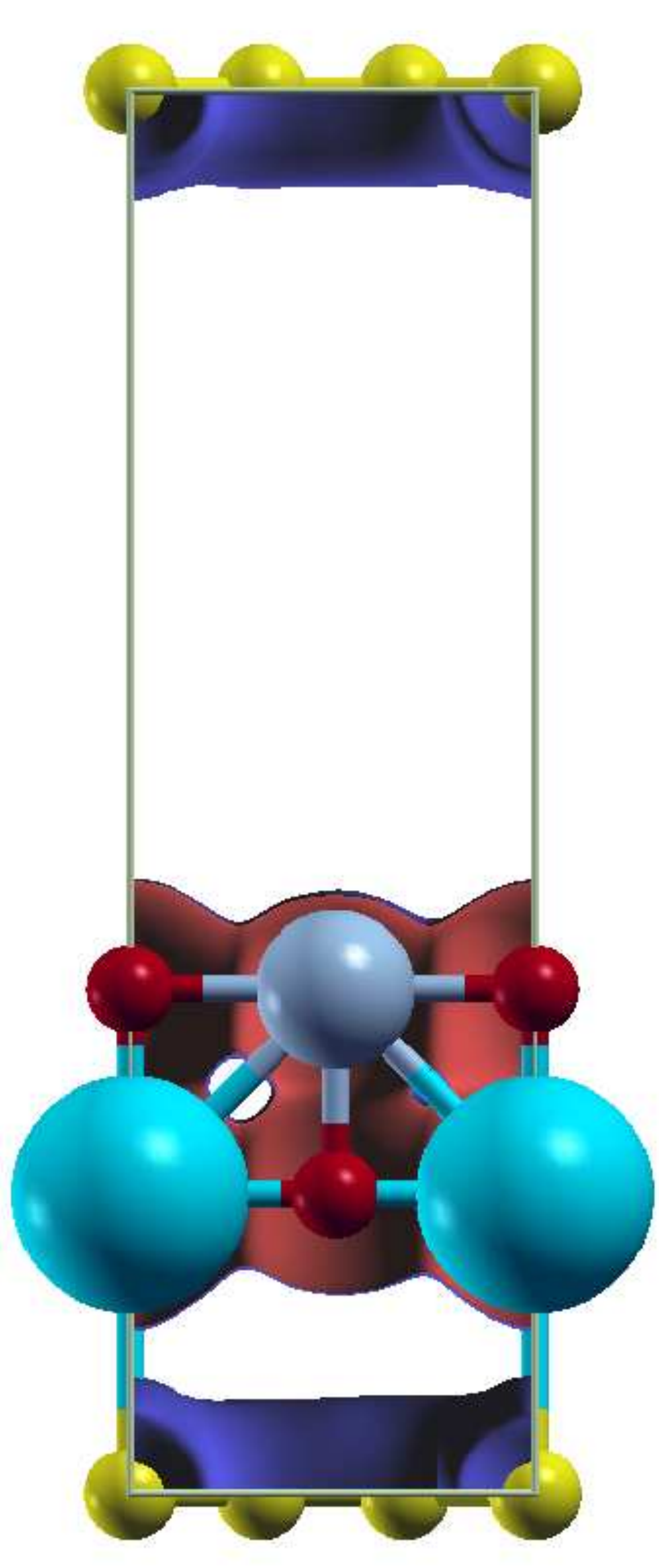}}
\subfigure[]{\includegraphics[width=4cm]{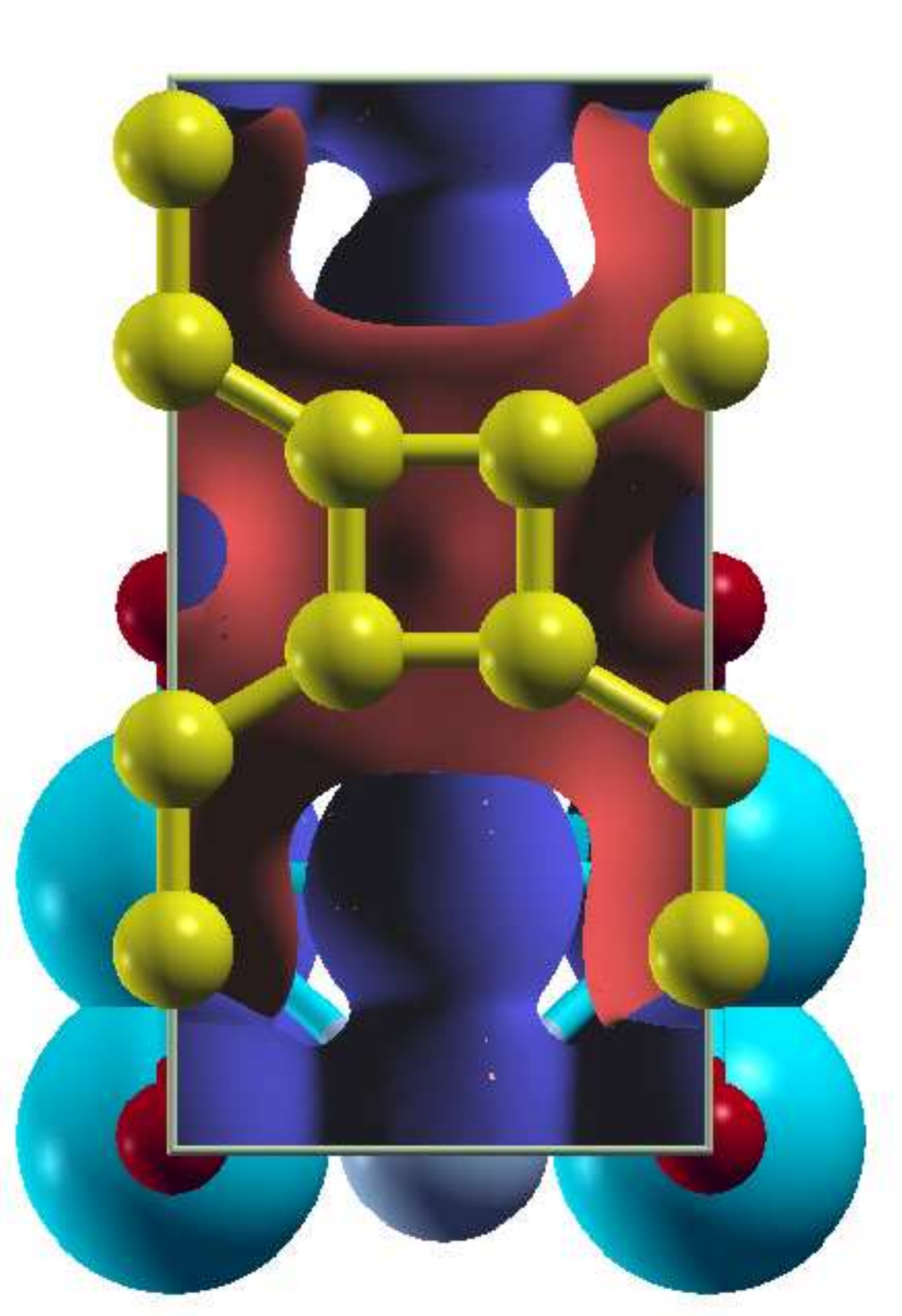}}
\caption{(a) Side and (b) top views of isosurface for the interface. Red regions represent charge accumulation, and blue regions represent charge depletion. The blue atoms are Sr, the red atoms are O, the gray atoms are Co, and the yellow atoms are C. The isovalue used was 0.002 electrons$/$Bohr$^3$.}
\label{isosurface}
\end{figure}

The charge transfer between graphene and SrCoO$_3$ was calculated in order to determine the areas of charge accumulation and charge depletion. To calculate the charge transfer, the converged charge density of the interface, pure SrCoO$_3$, and pure graphene were first obtained using the total energy calculations. The charge density of pure SrCoO$_3$ and pure graphene were then subtracted from the charge density of the interface to obtain the charge transfer. The resulting isosurface is displayed in Figure \ref{isosurface}, and the isovalue used was 0.002 electrons$/$Bohr$^3$. As shown in the figure, red regions indicate charge accumulation and blue areas indicate charge depletion. \par

The charge transfer could be caused by the strong hybridization of the C $2p$, Sr $3d$, Co $3d$, and O $2p$ orbitals. Figure \ref{isosurface} (a) clearly shows that the oxygen atoms facilitate the charge transfer through the interface. Charge is transported away from the carbon atoms towards SrCoO$_3$ through the oxygen atoms. These results are similar to those obtained by Hussain et al \cite{hussain2019} and Xiong et al \cite{xiong2013}, where the oxygen exhibits weak accumulation of charge and acts as a conducting channel. Figure \ref{isosurface} (b) indicates that there is also accumulation of charge on the graphene layer, confirming the conductivity of the interface as a whole. It has been suggested by Lang et al\cite{lang2017} that SrCoO$_3$ has the potential to prevent the agglomeration of graphene sheets through homogeneous dispersion. Due to this, it is possible that graphene not only increases the conductivity of the interface as a whole, but also can provide a buffer to accommodate the expansion and contraction of the cluster structure during charge/discharge cycles\cite{hussain2019}. \par
\begin{figure}
\subfigure[]{\includegraphics[width=7cm]{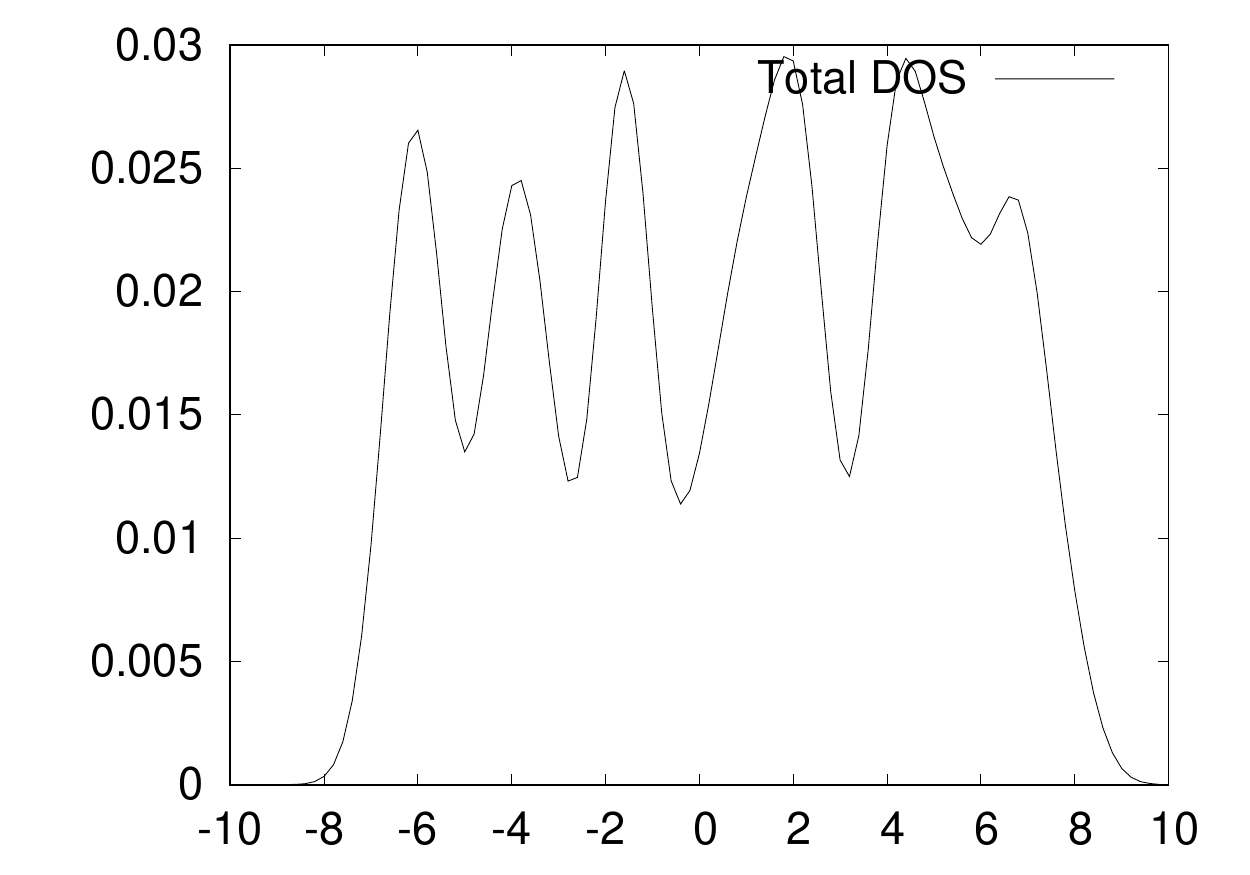}}
\subfigure[]{\includegraphics[width=7cm]{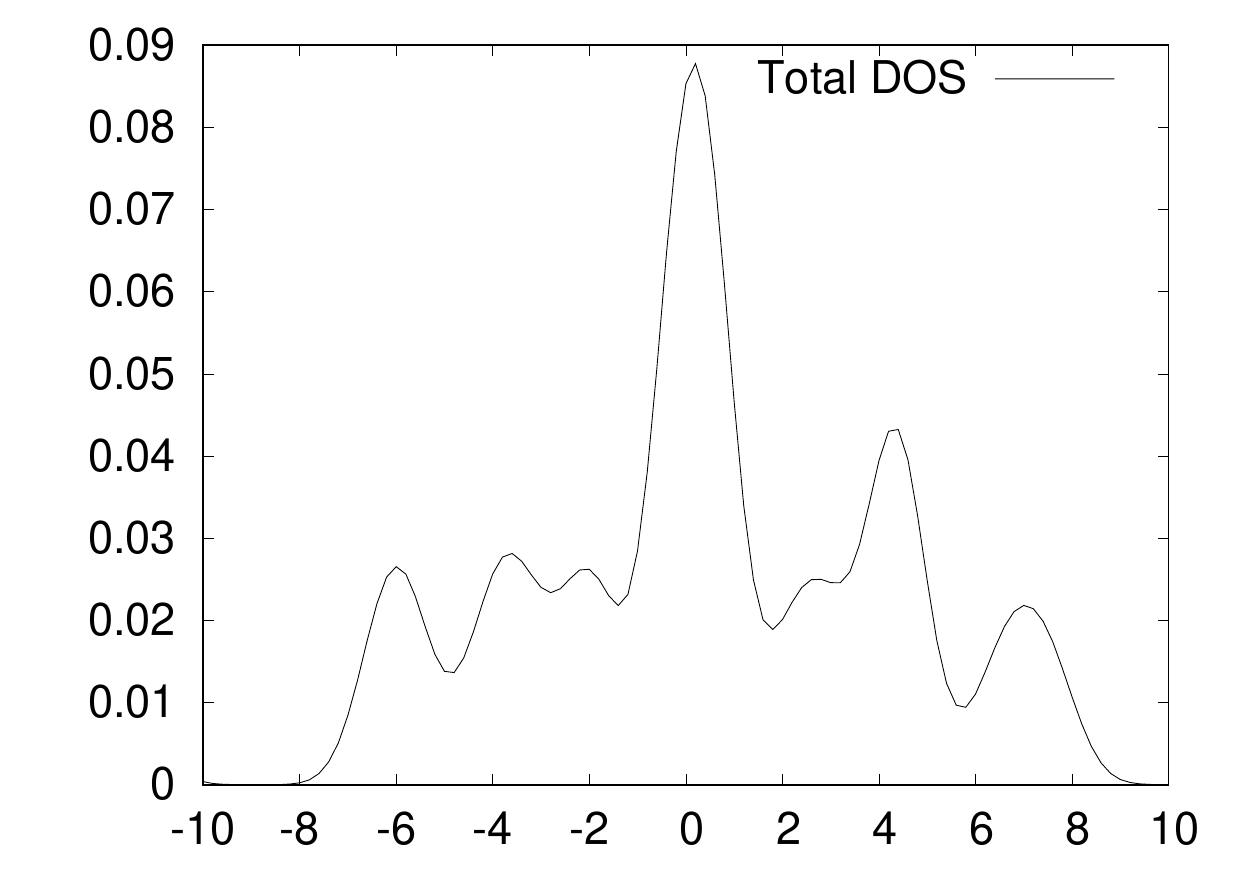}}
\subfigure[]{\includegraphics[width=7cm]{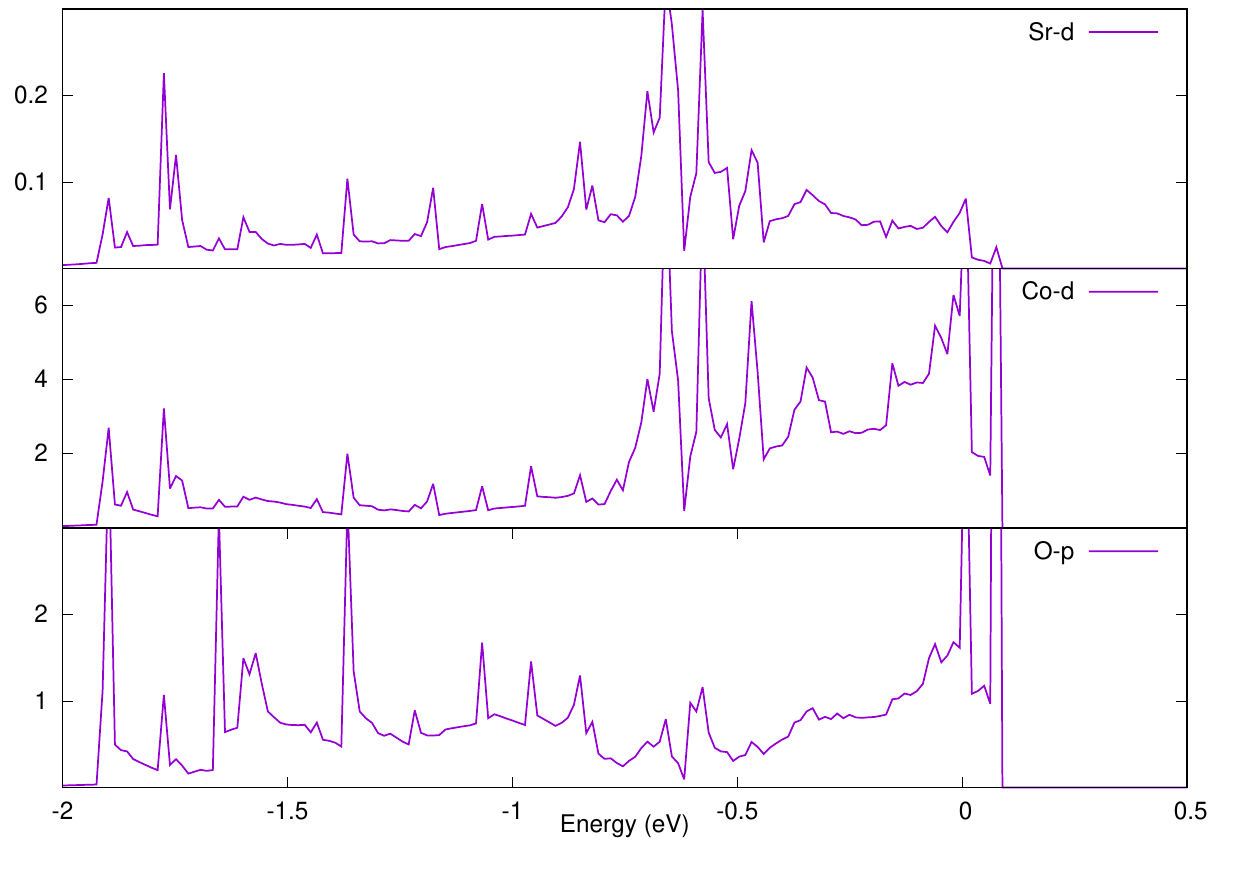}}
\subfigure[]{\includegraphics[width=7cm]{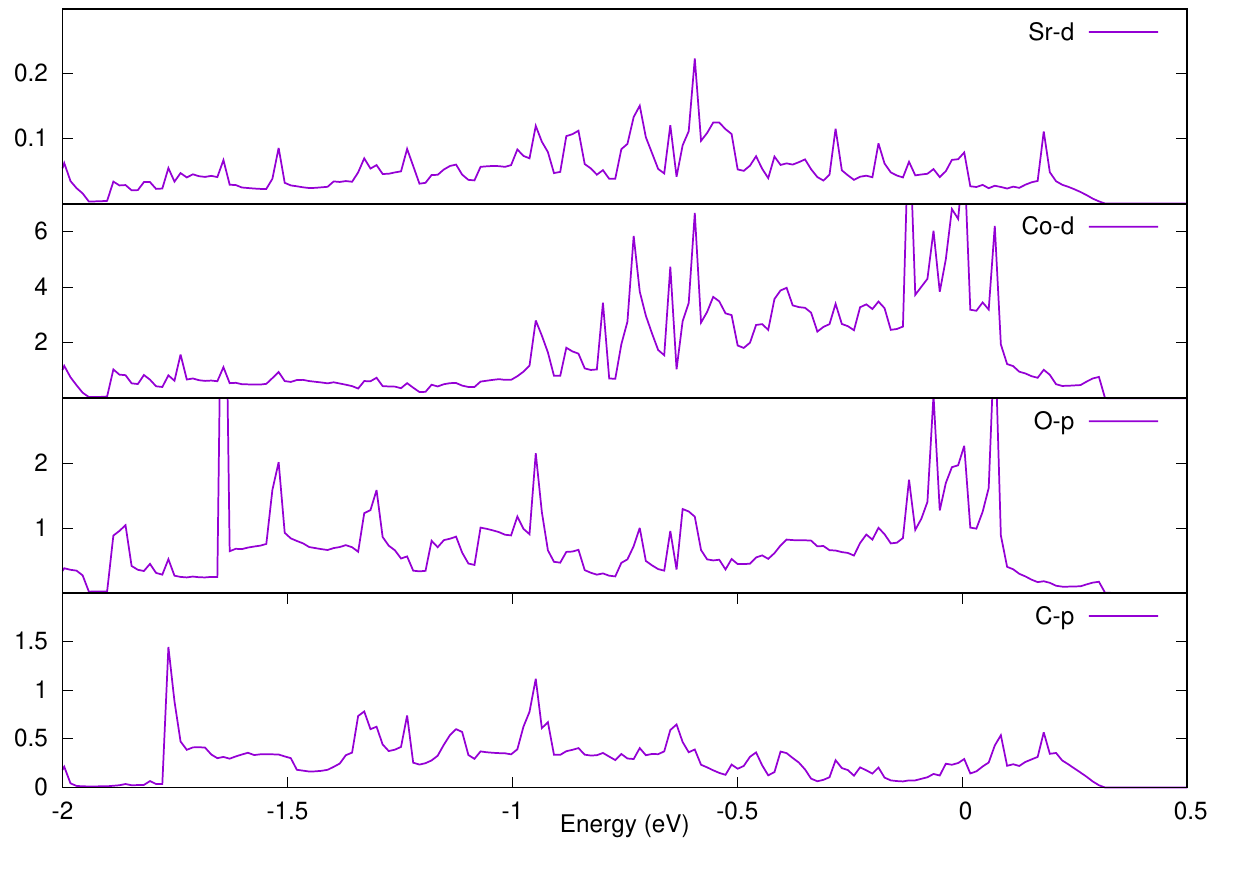}}
        \caption{TDOS of (a) pure SrCoO$_3$ and (b) the SrCoO$_3$$/$graphene interface. (c) shows the PDOS of pure SrCoO$_3$ and (d) shows the PDOS of the SrCoO$_3$$/$graphene interface. The increased number of states near the Fermi level for the interface in (b) and (d) indicates that it is extremely conductive due to the hybridization of the Sr $3d$, Co $3d$, O $2p$, and C $2p$.}
\label{interface_dos}
\end{figure}

The total DOS (TDOS) and projected DOS (PDOS) were calculated for the SrCoO$_3$$/$graphene interface to analyze the contribution of the Sr $3d$, Co $3d$, O $2p$, and C $2p$ orbitals to the conduction process. The TDOS of pure SrCoO$_3$ is displayed in Figure \ref{interface_dos} (a), and the TDOS of the SrCoO$_3$$/$graphene interface can be found in Figure \ref{interface_dos} (b). Compared to the TDOS of the pure perovskite, the interface calculation has more peaks in both the conduction and valence bands, indicating that the interface is more conductive overall than pure SrCoO$_3$. Figure \ref{interface_dos} (b) shows no band gap, which is to be expected because of the metallic nature of the SrCoO$_3$$/$graphene interface. Furthermore, the non-zero states at and around the Fermi level indicate an overlap between the valence and conduction bands and show that the states are shifting away from the valence bands towards the conduction bands. \par

The PDOS of the interface was calculated and is displayed in Figure \ref{interface_dos} (d), and the PDOS of the pure SrCoO$_3$ is shown in Figure \ref{interface_dos} (c) for comparison. The increased presence of non-zero localized states at and around the Fermi level indicates the increased conductivity of this material. It is clearly evident that the PDOS plot has many sharp peaks and that peaks are present for every material used in the interface, so it can be said that SrCoO$_3$ and graphene contribute equally to the conductivity of the interface. The increased number of coinciding peaks indicates the strong hybridization between C $2p$, O $2p$, Co $3d$, and Sr $3d$. The PDOS of the composite system is similar to the DOS of Sr, Co, and O as plotted by earlier studies\cite{van2001bulk}. Therefore, we can conclude that due to the relative stability and extremely high conductivity of this SrCoO$_3$$/$graphene interface, it has the potential to be an effective supercapacitor electrode. \par

\subsection{Oxygen vacancies}
A multitude of studies have concluded that inducing oxygen vacancies improves the conductivity of materials for supercapacitor applications\cite{yang2018, xiang2017two, wang2014}. Cheng et al\cite{cheng2017} specifically found that inducing oxygen vacancies in Co$_3$O$_4$ increased the specific capacitance of the complex material significantly. In the present study, we simulate one oxygen vacancy by removing one oxygen atom in a 2 $\times$ 2 $\times$ 1 supercell; similarly, we induce two oxygen vacancies by taking out one oxygen and its nearest neighboring oxygen atom. We then calculate the formation energy, band structure, and PDOS of the supercell with one V$_O$ and two V$_O$ and compare the two with pure SrCoO$_3$. \par

Oxygen vacancies were first induced in a pure SrCoO$_3$ supercell of size 2 $\times$ 2 $\times$ 1. After the oxygen was removed, the structure was relaxed again and the band structure and the PDOS were calculated. To simulate two oxygen vacancies, an oxygen atom and its nearest neighbor were both taken out. Table \ref{tab:vacancy_lattice} shows that with two vacancies, the relaxed lattice constant increases slightly. \par
\begin{center}
    \begin{table}
        \caption{Relaxed lattice constants for SrCoO$_3$ supercell with oxygen vacancies} \label{tab:vacancy_lattice}
    \begin{tabular}{cccc}
\hline
\hline
Number of vacancies & a (Bohr) & b (Bohr) & c (Bohr)\\
\hline
One & 14.4 & 14.4 & 7.22\\
Two & 14.5 & 14.5 & 7.23\\
\hline
\hline
\end{tabular}
    \end{table}
\end{center}

The defect formation energy of the complex structures was calculated. To do this, the total energy of pure SrCoO$_3$ was determined using the relaxed lattice constants in Table \ref{interface_structure} and converged kinetic energy cutoff and Monkhorst-Pack k-point grids displayed in Table \ref{tab:ecutkpt}. The formalism defined in Eq. (\ref{eqn:vacancy_energy}) was then used to determine the defect formation energy of the oxygen vacancies. The formation energy of the complex with one oxygen vacancy is 2.75 eV, while the formation energy of the complex with two oxygen vacancies is 7.62 eV. Since a lower formation energy is indicative of a more stable compound, the complex with one oxygen vacancy is more stable than the complex with two oxygen vacancies. This is to be expected, as inducing vacancies disrupts the crystal structure and therefore decreases the stability of the system. Compared to other studies which have analyzed oxygen vacancies with transition metal oxides, the complex with one oxygen vacancy is relatively stable. Previous studies have shown that inducing an oxygen vacancy in TiO$_2$ results in a formation energy of around 2-5 eV \cite{na2006}. Since the formation energy of one oxygen vacancy is similar to this range, we conclude that the formation energies of these complexes are reasonable. \par

\begin{figure}
\subfigure[]{\includegraphics [width=7cm]{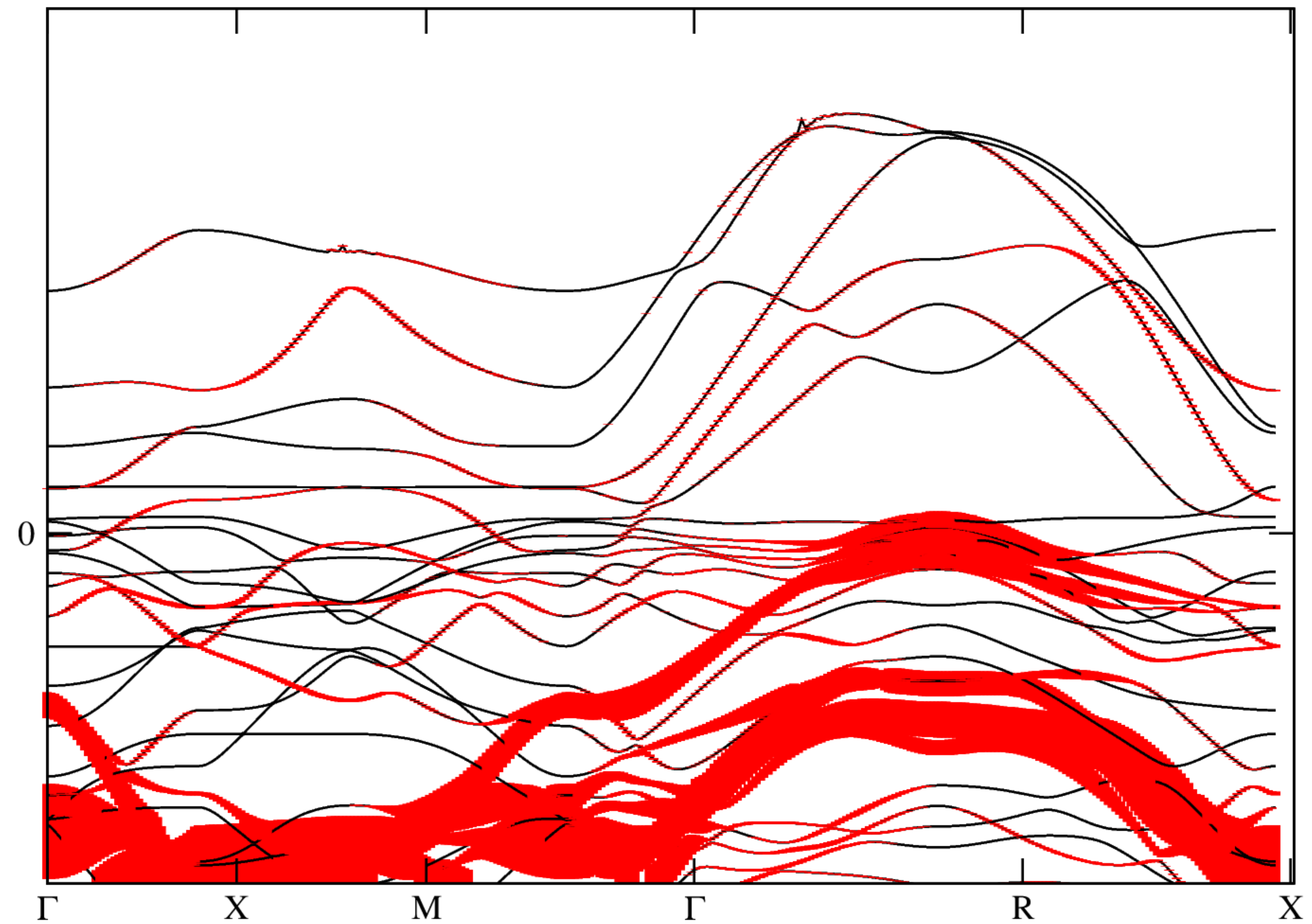}}
\subfigure[]{\includegraphics [width=7cm]{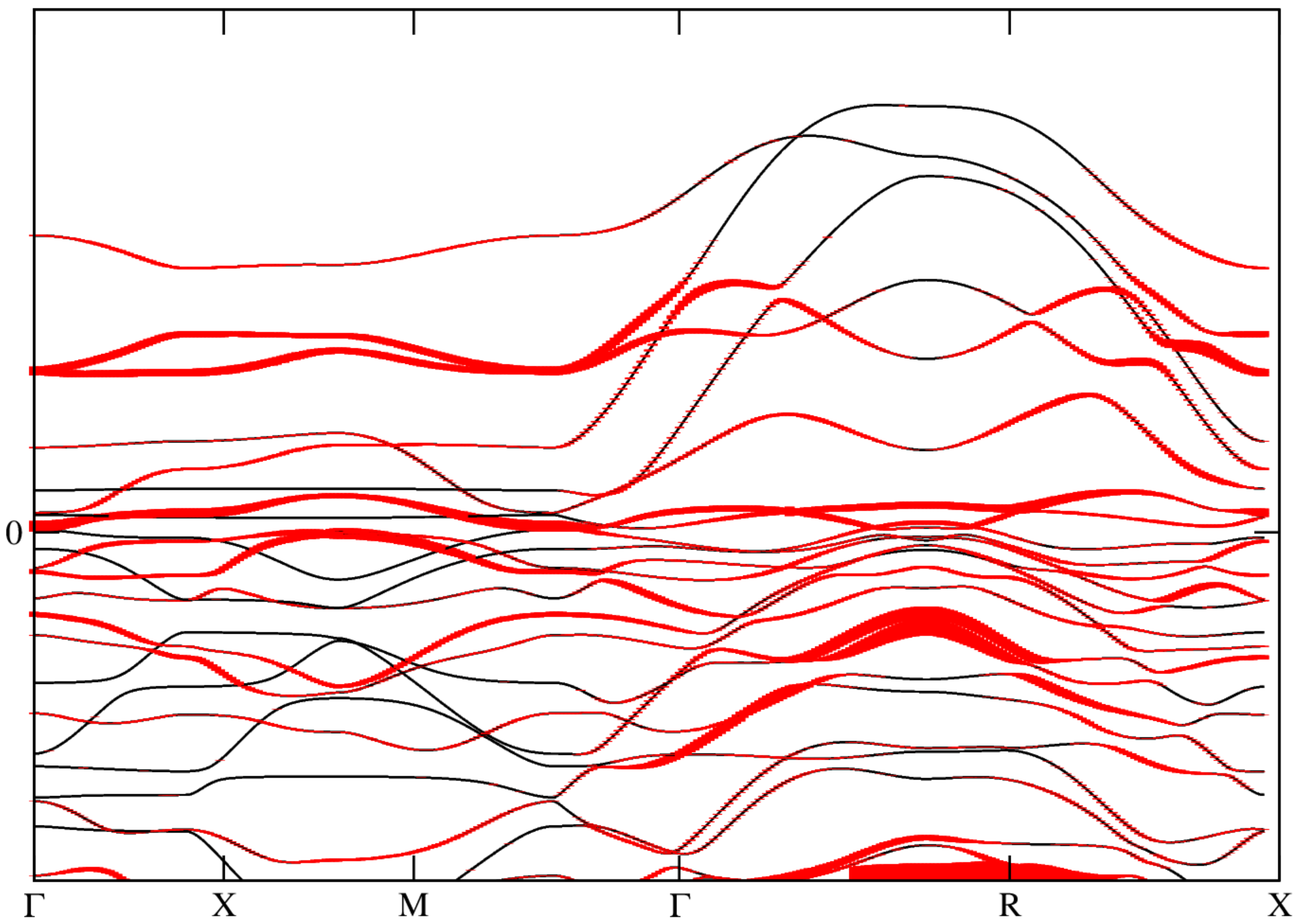}}
    \caption{Fat-band structure for an SrCoO$_3$ complex with (a) one oxygen vacancy and (b) two oxygen vacancies. The contributions of the oxygen $2p$ orbital are shown in red. The high symmetry k-points for the cubic structure were used and are: $\Gamma$ (0, 0, 0), X(0, 0.5, 0), M(0.5, 0.5, 0.0), and R(0.5, 0.5, 0.5)}
\label{vacancy_bands}
\end{figure}

The fat-band structures of the one-vacancy complex and the two-vacancy complex were calculated and are displayed in Figure \ref{vacancy_bands}. The contributions by the oxygen $2p$ orbital are plotted in red, and it is evident that inducing oxygen vacancies increases the contributions of oxygen to the bands around the Fermi level. The oxygen $2p$ orbital contributes more to the two-vacancy complex than to the one-vacancy complex, demonstrating that the contributions of the oxygen $2p$ orbital increase as the number of vacancies increases. As expected, there is little to no band gap near the Fermi level, which indicates that the materials in question are metallic. \par
\begin{figure}[hpt]
    \includegraphics [width=10cm]{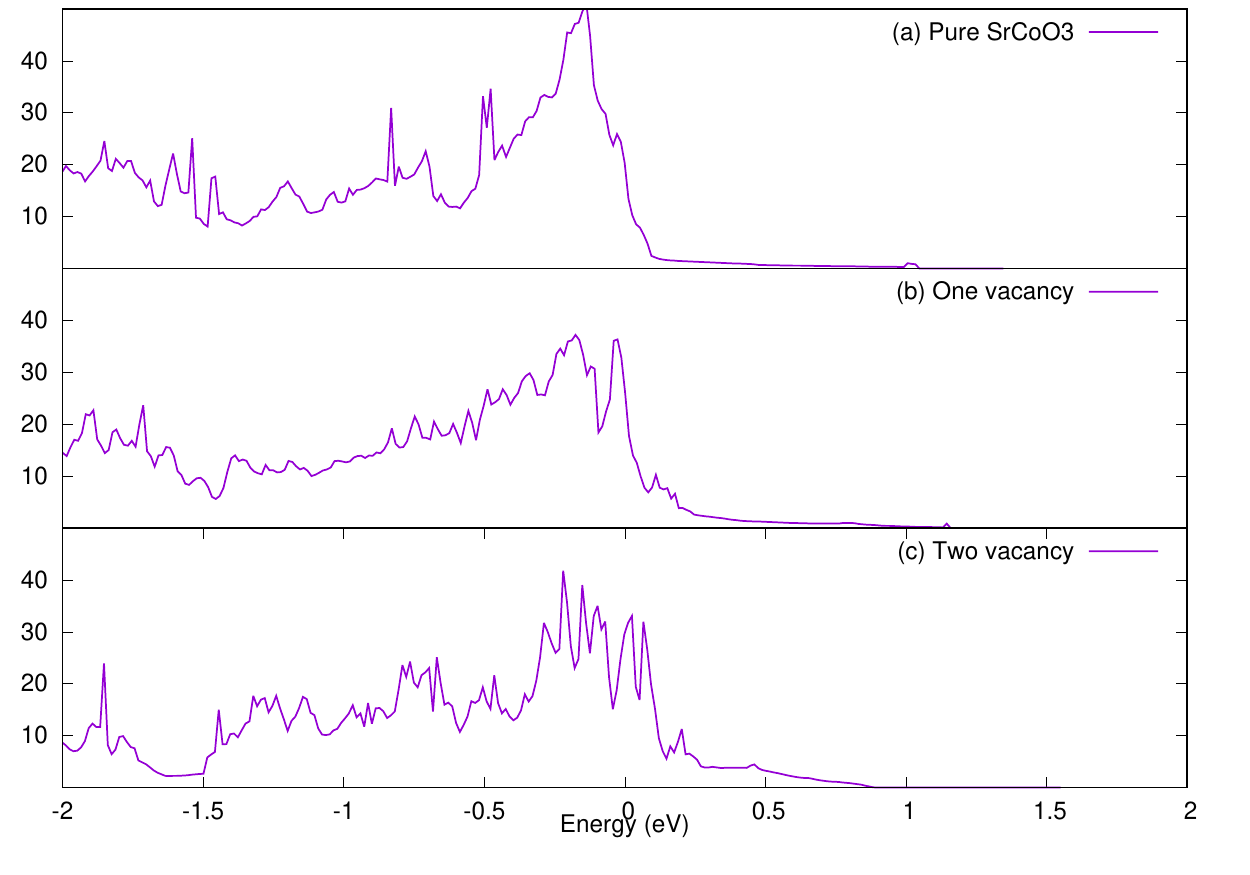}
        \caption{Total DOS for (a) pure SrCoO$_3$, (b) an SrCoO$_3$ complex with one vacancy, and (c) an SrCoO$_3$ complex with two vacancies. Since there is little to no band gap around the Fermi level, we can conclude that these materials are metallic.}
\label{vacancy_dos}
\end{figure}

In order to yield further insights into the conductivity of the complexes with vacancies, TDOS and PDOS were calculated. To calculate the DOS, the total energy was determined using the converged kinetic energy cutoffs and Monkhorst-Pack k-point grids, as well as the relaxed lattice vectors shown in Table \ref{tab:vacancy_lattice}. The TDOS of the complexes can be found in Figure \ref{vacancy_dos} (b) and (c). The total DOS of the pure SrCoO$_3$ is also shown in Figure \ref{vacancy_dos} (a) for comparison. It is clearly evident from \ref{vacancy_dos} (b) and \ref{vacancy_dos} (c) that inducing oxygen vacancies improves the conductivity of the complex because of the increased number of peaks in the TDOS in figures \ref{vacancy_dos}(b) and (c). Figures \ref{vacancy_dos} (b) and (c) show no band gap, which is to be expected because of the metallic nature of SrCoO$_3$. In addition, the non-zero states at and around the Fermi level indicate an overlap between the valence and conduction bands and show that the states are shifting away from the valence bands towards the conduction bands.

\begin{figure}[hpt]
\subfigure[]{\includegraphics [width=7cm]{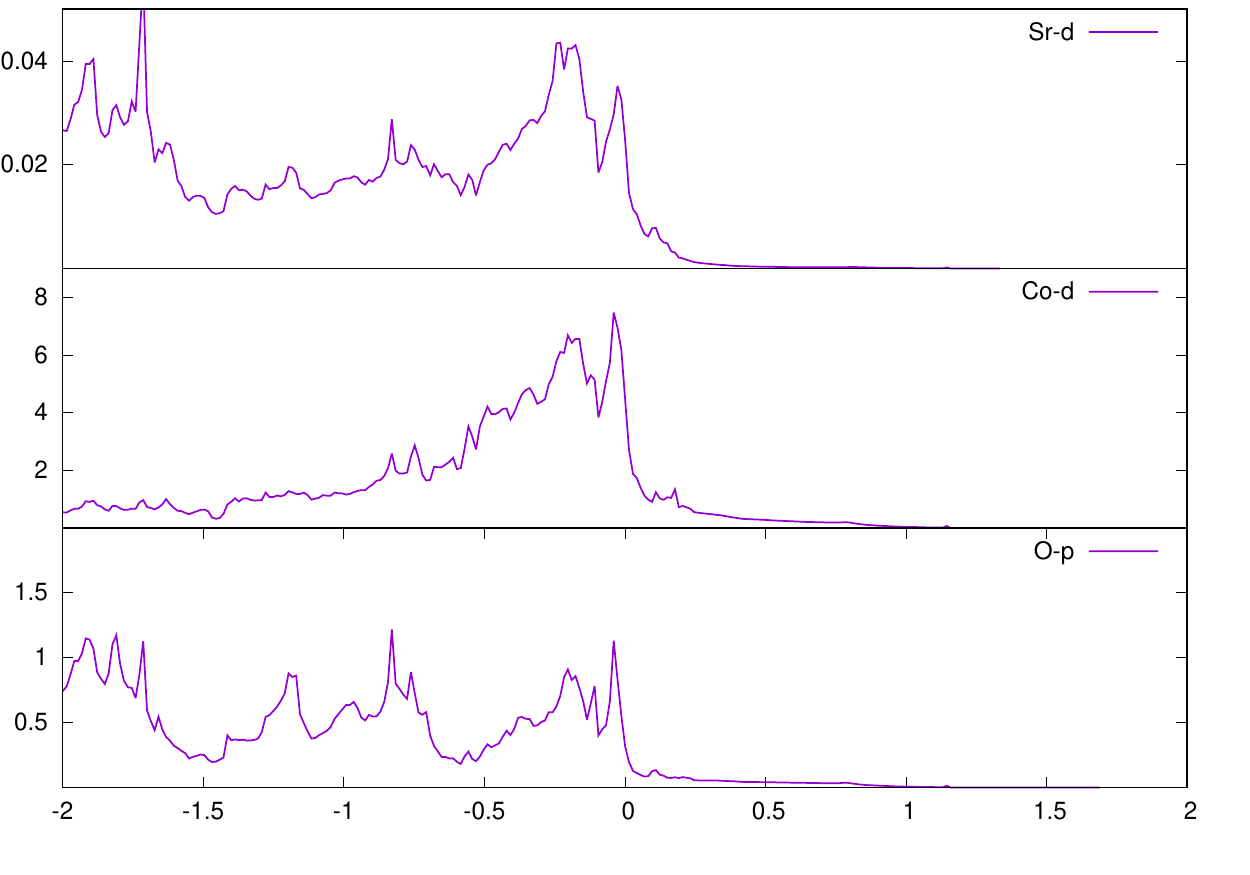}}
\subfigure[]{\includegraphics [width=7cm]{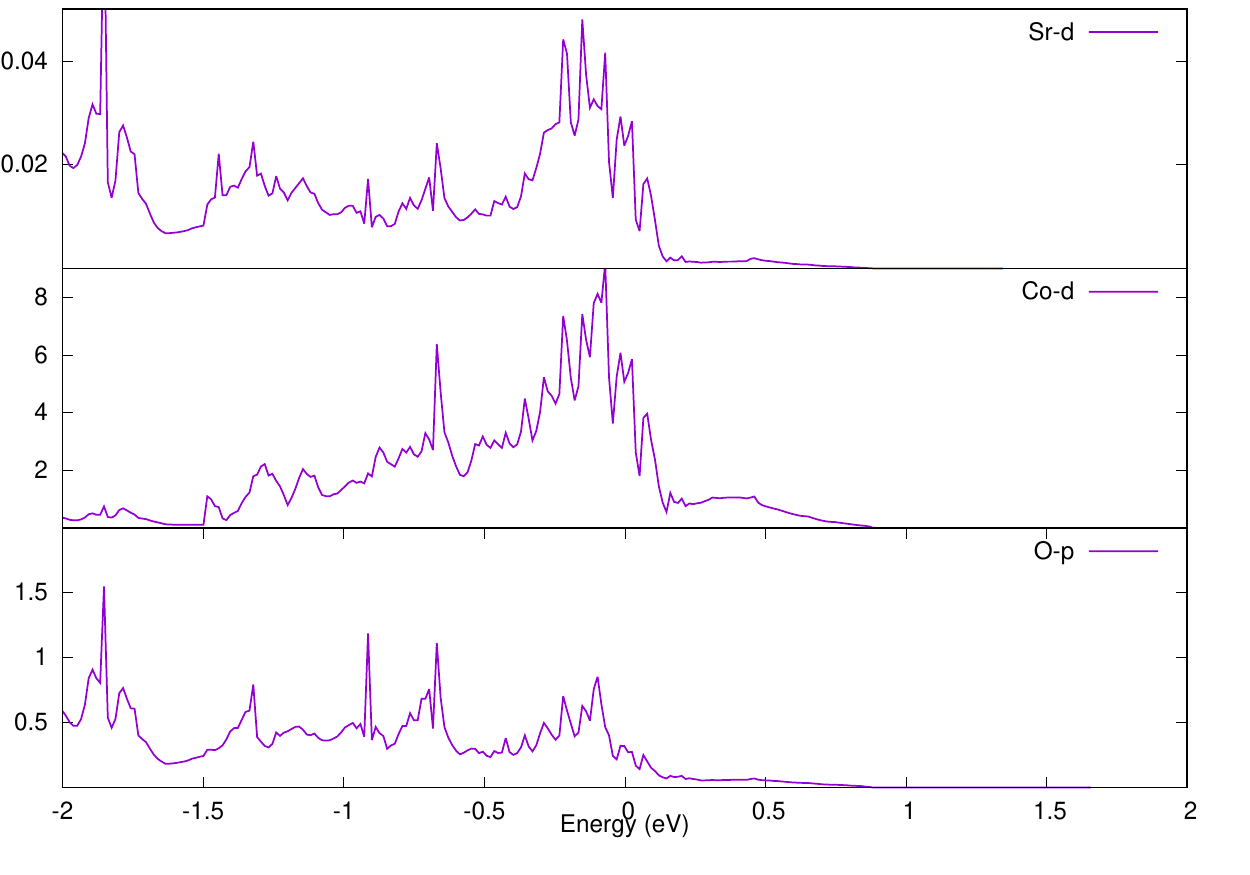}}
\caption{PDOS for an SrCoO$_3$ supercell with (a) one oxygen vacancy and (b) two oxygen vacancies. The PDOS of the Sr $2p$ orbital is shown at the top, then the Co $3d$ orbital, and then O $2p$ orbital is displayed.}
\label{vacancy_pdos}
\end{figure}

The PDOS of the complexes were also calculated in order to determine the contributions of Sr $3d$, Co $3d$, and O $2p$ with vacancies to the conduction process. The PDOS reveals that vacancies increase the conductivity of SrCoO$_3$ tremendously, and there is also a considerable increase in conductivity from one vacancy to two vacancies. Figures \ref{vacancy_pdos} (a) and (b) show that there are localized states at and around the Fermi level, indicating an increase in conductivity. In addition, peaks are present for all elements, indicating that Sr, Co, and O contribute equally to the conduction process. The non-zero states at the Fermi level show that there is an overlap between the valence and conduction band edges that reveals a shift away from the valence band (VB) toward conduction bands (CB). The increased number of states from one vacancy to two vacancies reveals the strong hybridization of the Sr $3d$, Co $3d$, and O $2p$ orbitals caused by the vacancy. Due to the increase in conductivity caused by vacancies and their relative stability, we conclude that inducing oxygen vacancies in SrCoO$_3$ increase its performance as a supercapacitor electrode. \par

\begin{figure}[hpt]
\subfigure[]{\includegraphics [width=5cm]{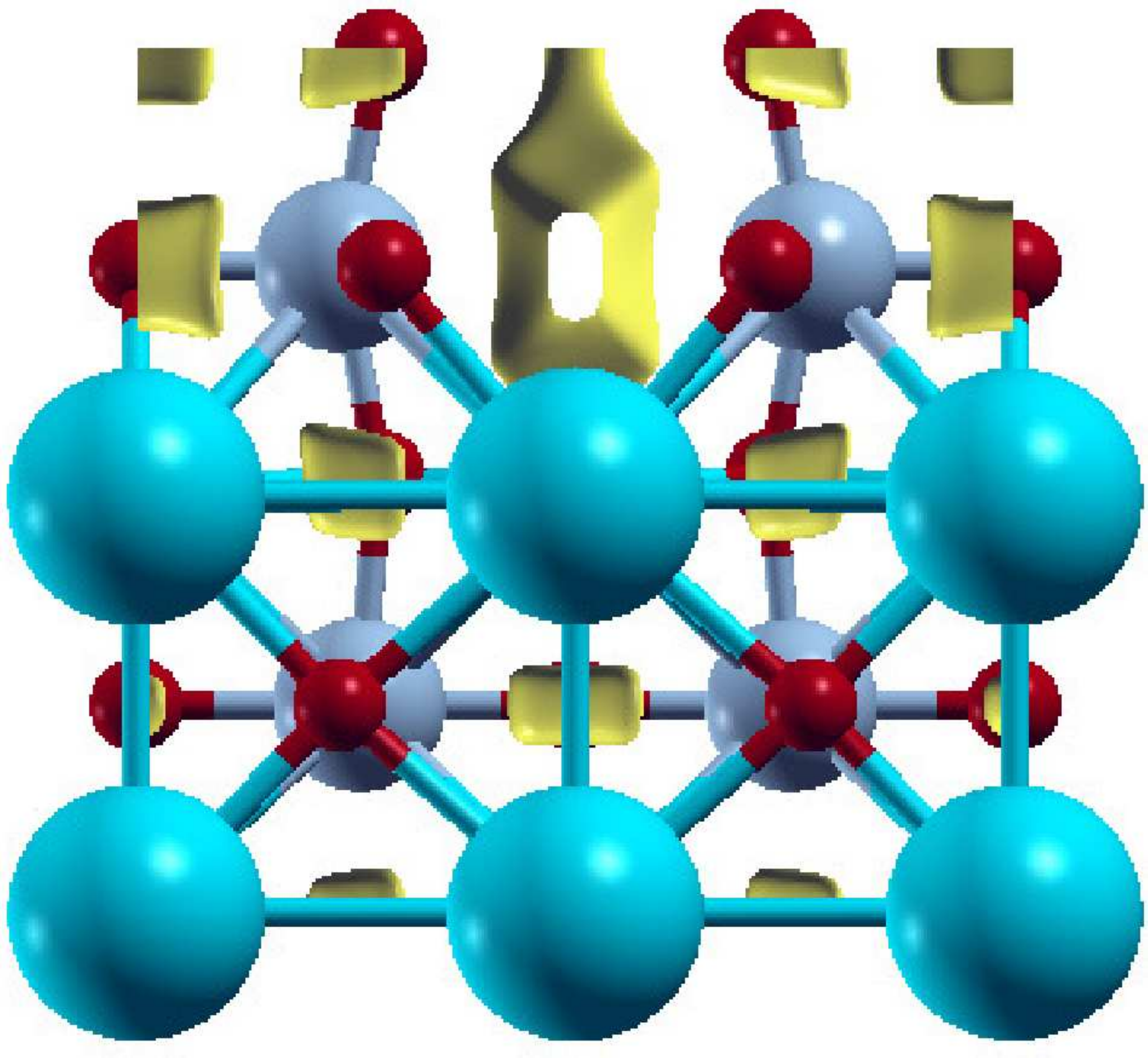}}
\subfigure[]{\includegraphics [width=5cm]{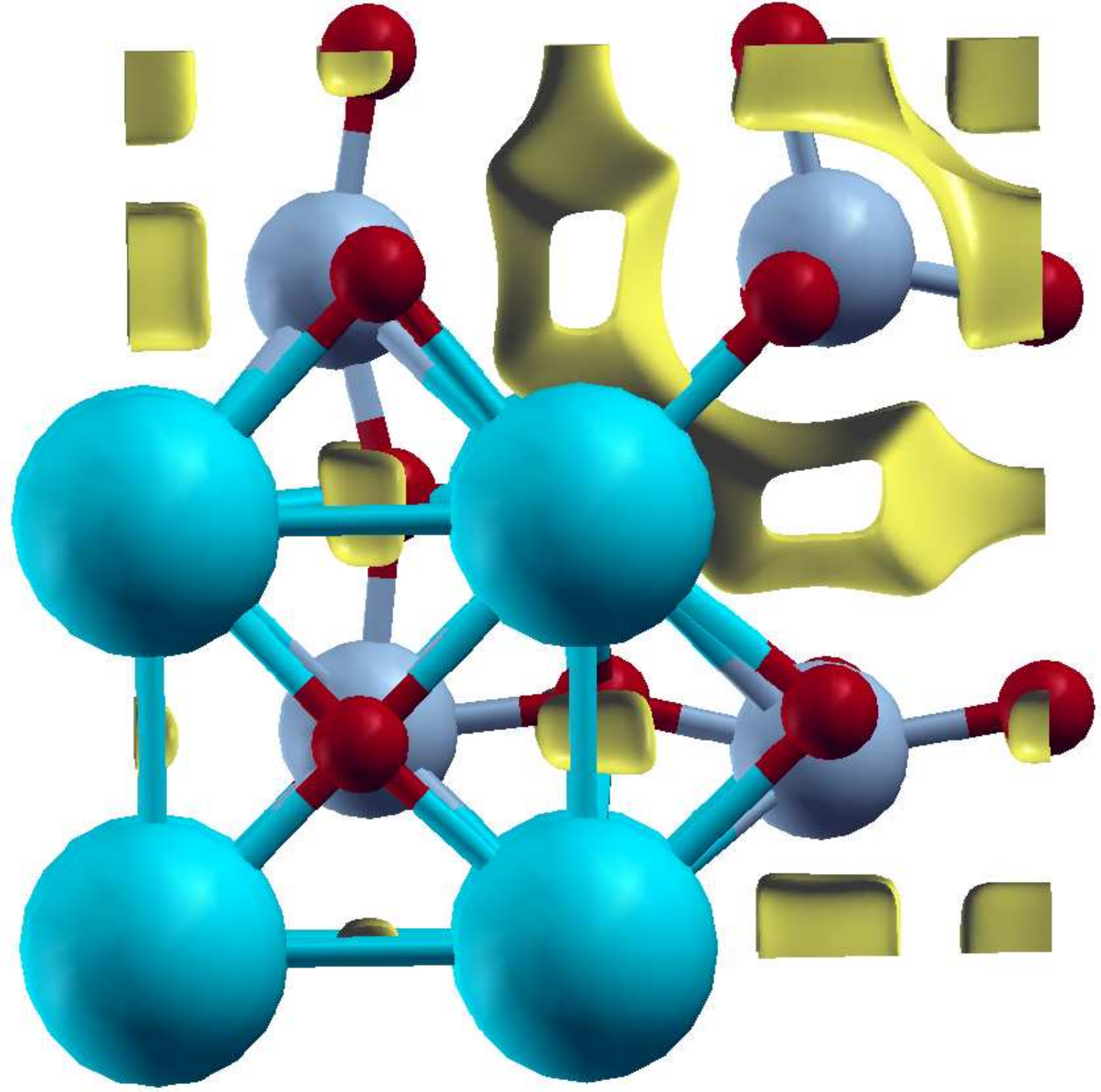}}
        \caption{Isosurface diagrams for SrCoO$_3$ complexes containing (a) one oxygen vacancy and (b) two oxygen vacancies.  The yellow areas indicate charge accumulation. The blue atoms are Sr, the red atoms are O, and the gray atoms represent Co.}
\label{one_vacancy_iso}
\end{figure}

The converged charge density was calculated for both one V$_O$ and 2 V$_O$ and the results are displayed in figure \ref{one_vacancy_iso}. The charge density was obtained by calculating the total energy of the complex using the kinetic energy cutoffs and k-point grids of pure SrCoO$_3$ and the relaxed lattice constants of the complex. The isovalue used was 0.03 electrons$/$Bohr$^3$. Figure \ref{one_vacancy_iso} (a) shows that the charge accumulation is centered around the vacancies, which is well in agreement with published studies \cite{yang2018}. Similarly, Figure \ref{one_vacancy_iso}(b) shows two concentrated areas of charge accumulation caused by the two vacancies in the structure. This accumulation is most likely caused by the increased electron delocalization of the surrounding Sr and Co atoms as a result of the oxygen vacancy. In addition, the charge accumulation could also have been caused by the strong hybridization of the Sr $3d$, Co $3d$, and O $2p$ orbitals, which supports the PDOS calculations shown in Figure \ref{vacancy_pdos}. \par

\subsection{Doping}
We choose Mo, V, P, and Nb as dopants to investigate how they increase the OER of SrCoO$_3$. We also study whether nonmetallic dopants can achieve the same level of conductivity and stability as the metallic dopants.\par

To achieve 25\% substitutional doping at the B-site of SrCoO$_3$, a 2 $\times$ 2 $\times$ 1 supercell was used, in which one cobalt atom was replaced with the dopant. The complex was then relaxed and the band structure, PDOS, and formation energy were calculated for each dopant and compared to determine the most effective dopant.  \par

\begin{figure}
\subfigure[]{\includegraphics [width=6cm]{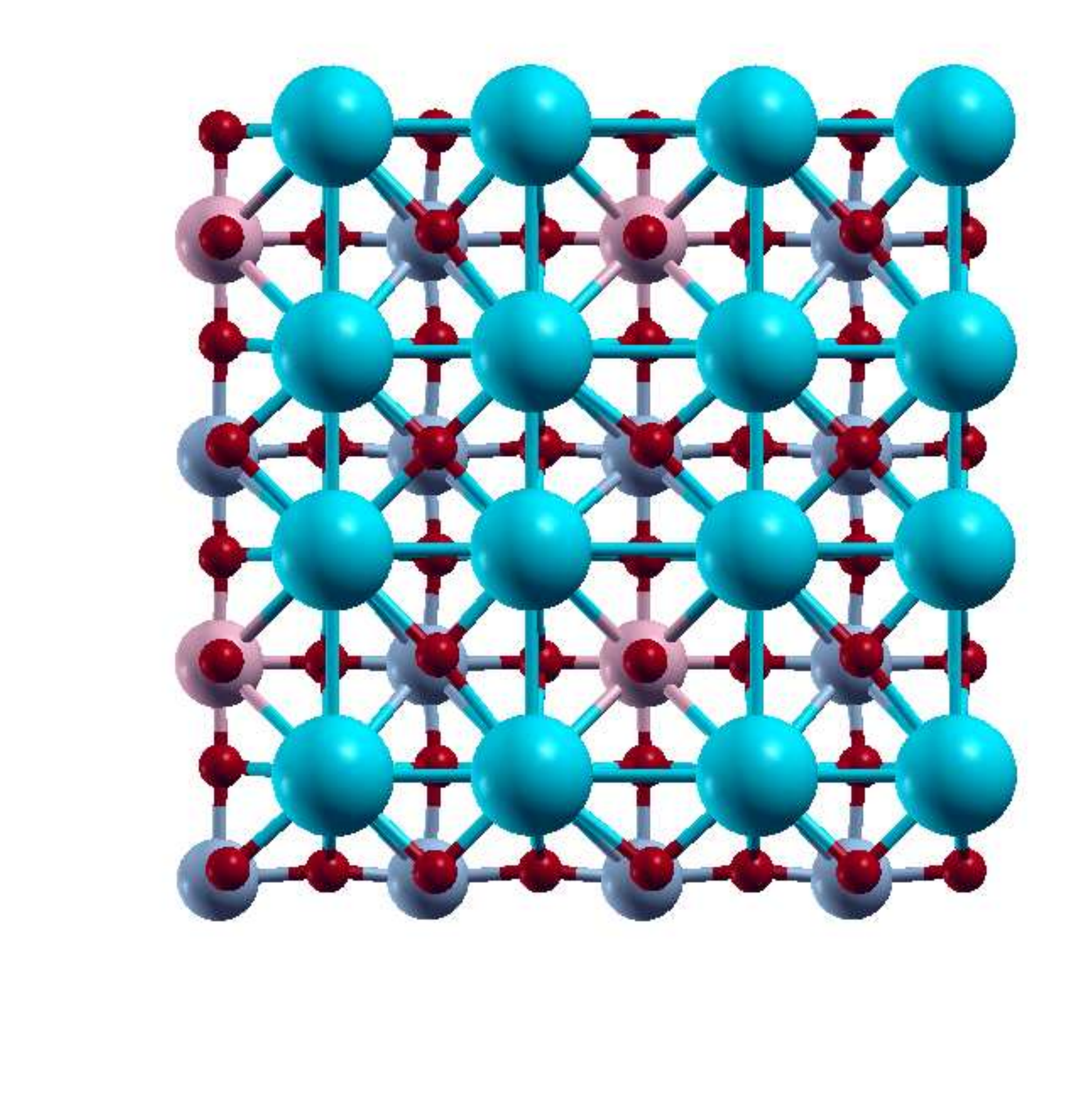}}
\subfigure[]{\includegraphics [width=6cm]{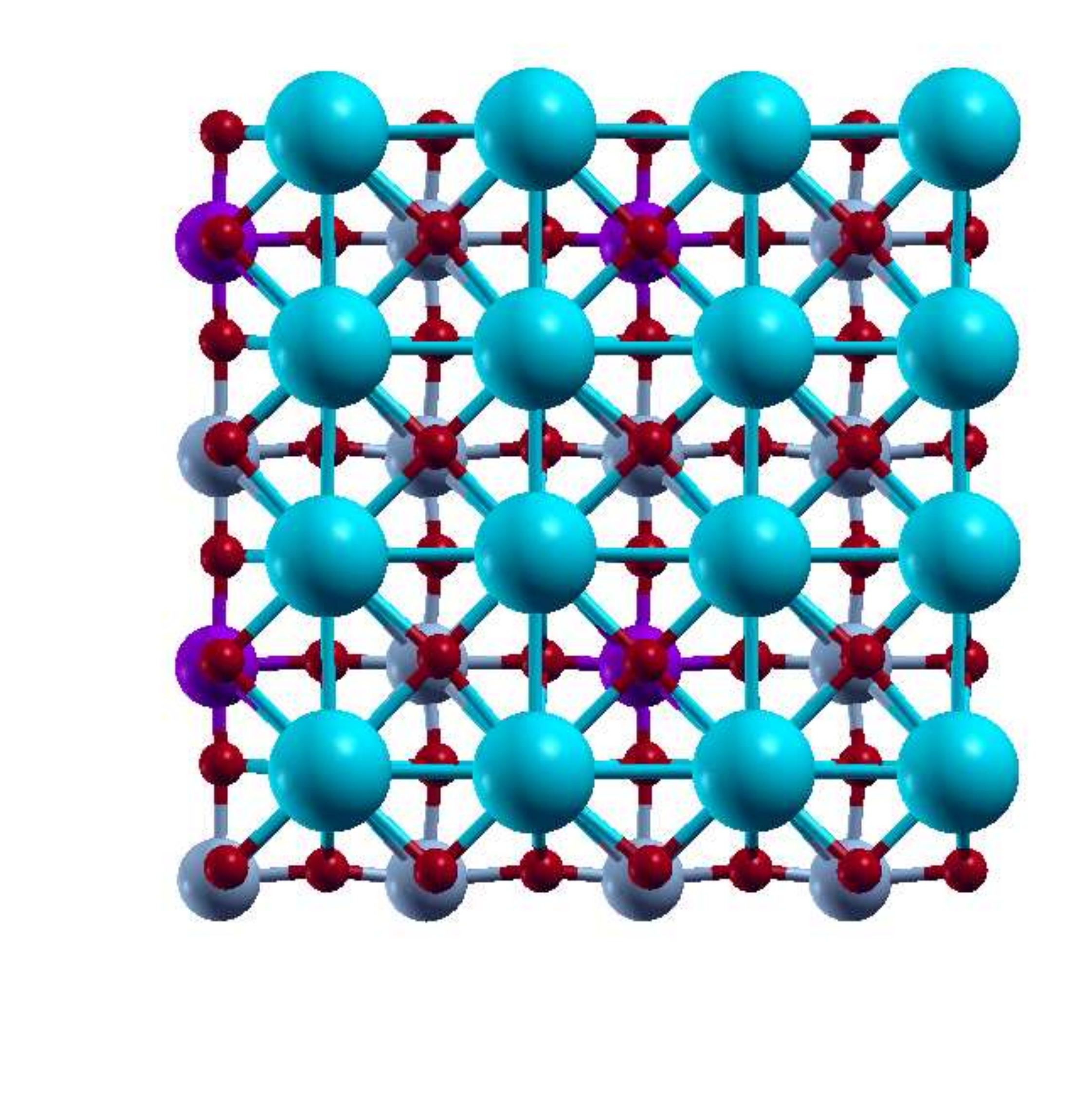}}
\subfigure[]{\includegraphics [width=6cm]{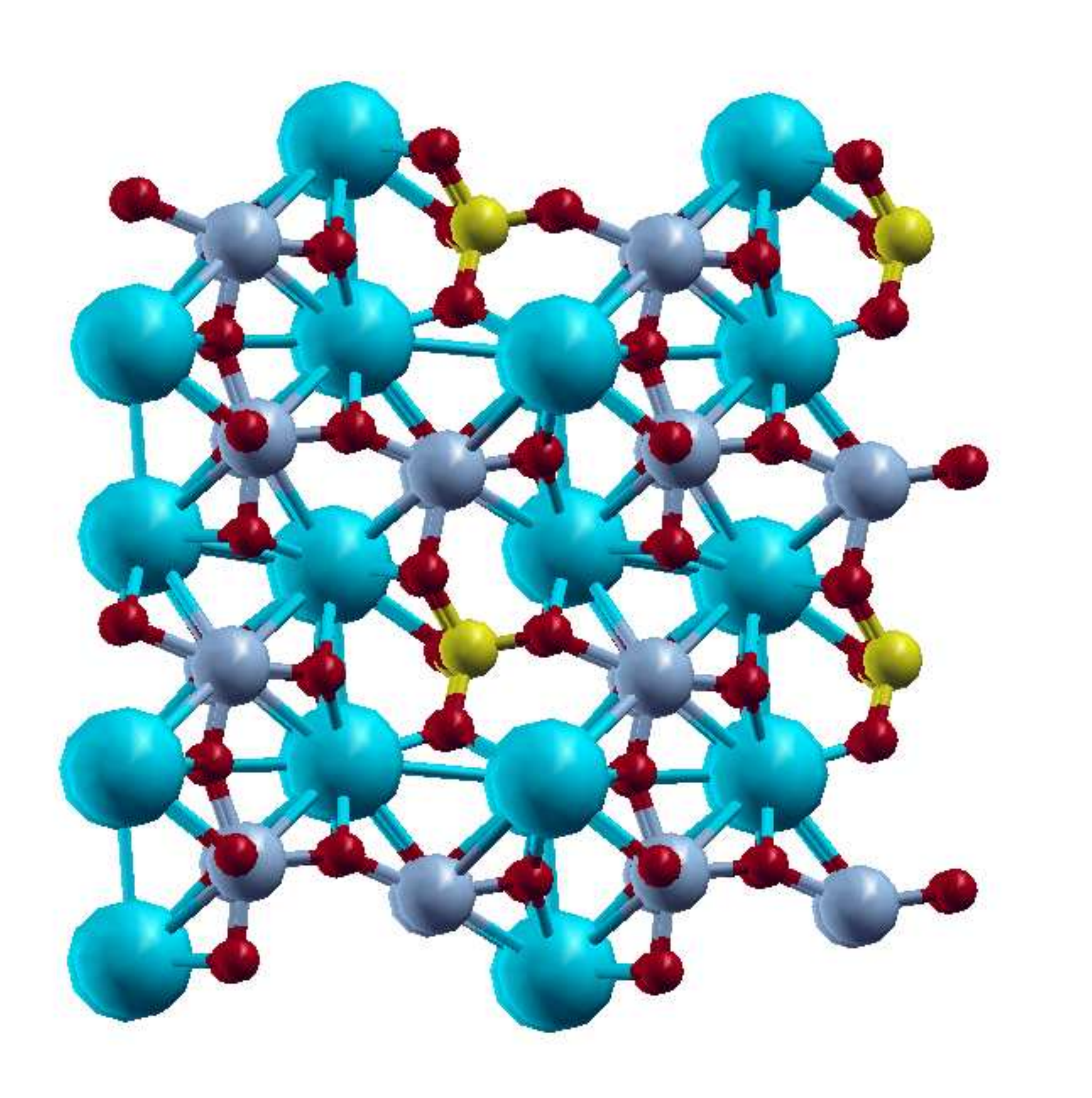}}
\subfigure[]{\includegraphics [width=6cm]{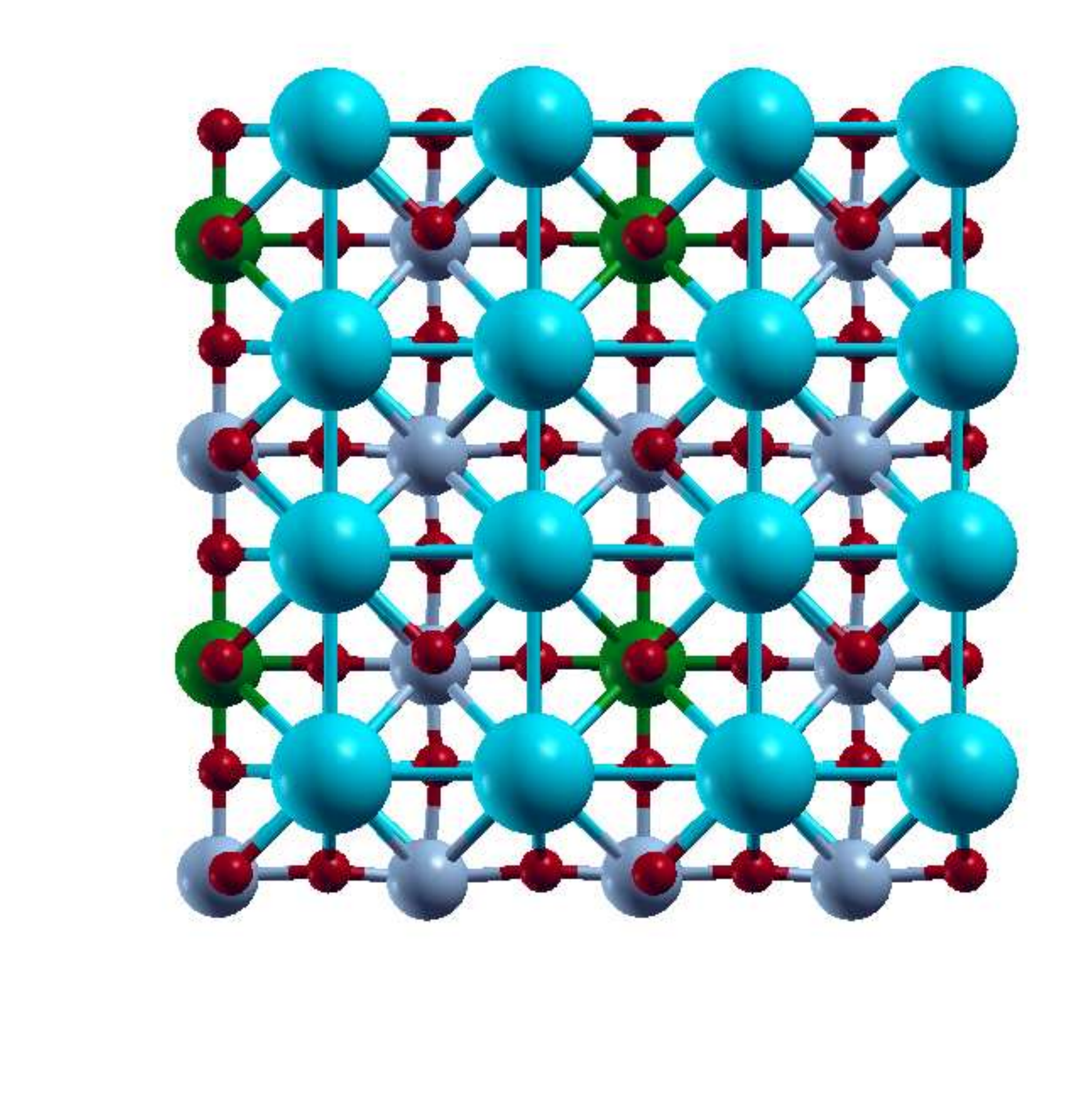}}
\caption{Atomic structure for (a) Mo-doped SrCoO$_3$, (b) V-doped SrCoO$_3$, (c) P-doped SrCoO$_3$, (d) Nb-doped SrCoO$_3$. The blue atoms represent Sr, the gray atoms represent Co, the red atoms represent O, the pink atoms represent Mo, the purple atoms represent V, the yellow atoms represent P, and the green atoms represent Nb.}
\label{dopant_atomic_structure}
\end{figure}

Once a cobalt atom was substituted for one of the dopants, the resulting structure was relaxed again. The fully relaxed structure of the complex remained cubic when the dopant was Mo, V, or Nb, which confirms experimental results, as these dopants do not alter the basic structure of SrCoO$_3$. For P-doped SrCoO$_3$, however, the structure was altered from cubic to tetragonal, which supports the experimental results found by Zhu et al \cite{zhu2016}. Table \ref{tab:dopant_relaxed_dim} displays the relaxed lattice constants of each complex. The atomic structure is displayed in Figure  \ref{dopant_atomic_structure}. \par
\begin{center}
\begin{table}
\caption{Relaxed lattice constants of doped SrCoO$_3$} \label{tab:dopant_relaxed_dim}
\begin{tabular}{cccc}
\hline
\hline
Compound & a (Bohr) & b (Bohr) & c (Bohr) \\
\hline
Mo-doped SrCoO$_3$ & 14.64 & 14.64 &  7.32 \\
V-doped SrCoO$_3$ & 14.52 & 14.52 & 7.26 \\
P-doped SrCoO$_3$ & 14.77 & 14.77 & 7.39 \\
Nb-doped SrCoO$_3$ & 14.69 & 14.69 & 7.34 \\
\hline
\hline
\end{tabular}
\end{table}
\end{center}

The formation energy for all dopants was calculated. To do this, the total energy was calculated for the doped complex, pure SrCoO$_3$, pure cobalt, and the pure dopant using the converged kinetic energy cutoffs, Monkhorst-Pack k-point grids, and relaxed lattice vectors. The formalism displayed in Eq. (\ref{eqn:dopant_energy}) was used to calculate the formation energy of the complex, and the results are displayed in Table \ref{tab:dopant_energy}. It has been well-documented by Hussain et al\cite{hussain2018} and others that a lower formation energy is an indicator of stability of a structure. These DFT calculations reveal that niobium-doped SrCoO$_3$ has the lowest formation energy of -9.89 eV, indicating that it is the most stable compound of the four dopants. All of the formation energies are negative, which indicates that exothermic reactions are taking place. This further confirms that all of the complexes are stable as they do not gain heat during the doping reaction. \par

\begin{center}
\begin{table}
\caption{Formation energy for doped SrCoO$_3$ with different dopants}
\label{tab:dopant_energy}
\begin{tabular}{cc}
\hline
\hline
Compound & Formation Energy (eV) \\
\hline
Mo-doped SrCoO3 & -7.46 \\
V-doped SrCoO3 & -7.88 \\
P-doped SrCoO3 & -6.07 \\
Nb-doped SrCoO3 & -9.89 \\
\hline
\hline
\end{tabular}
\end{table}
\end{center}

\begin{figure}
\subfigure[]{\includegraphics [width=6cm]{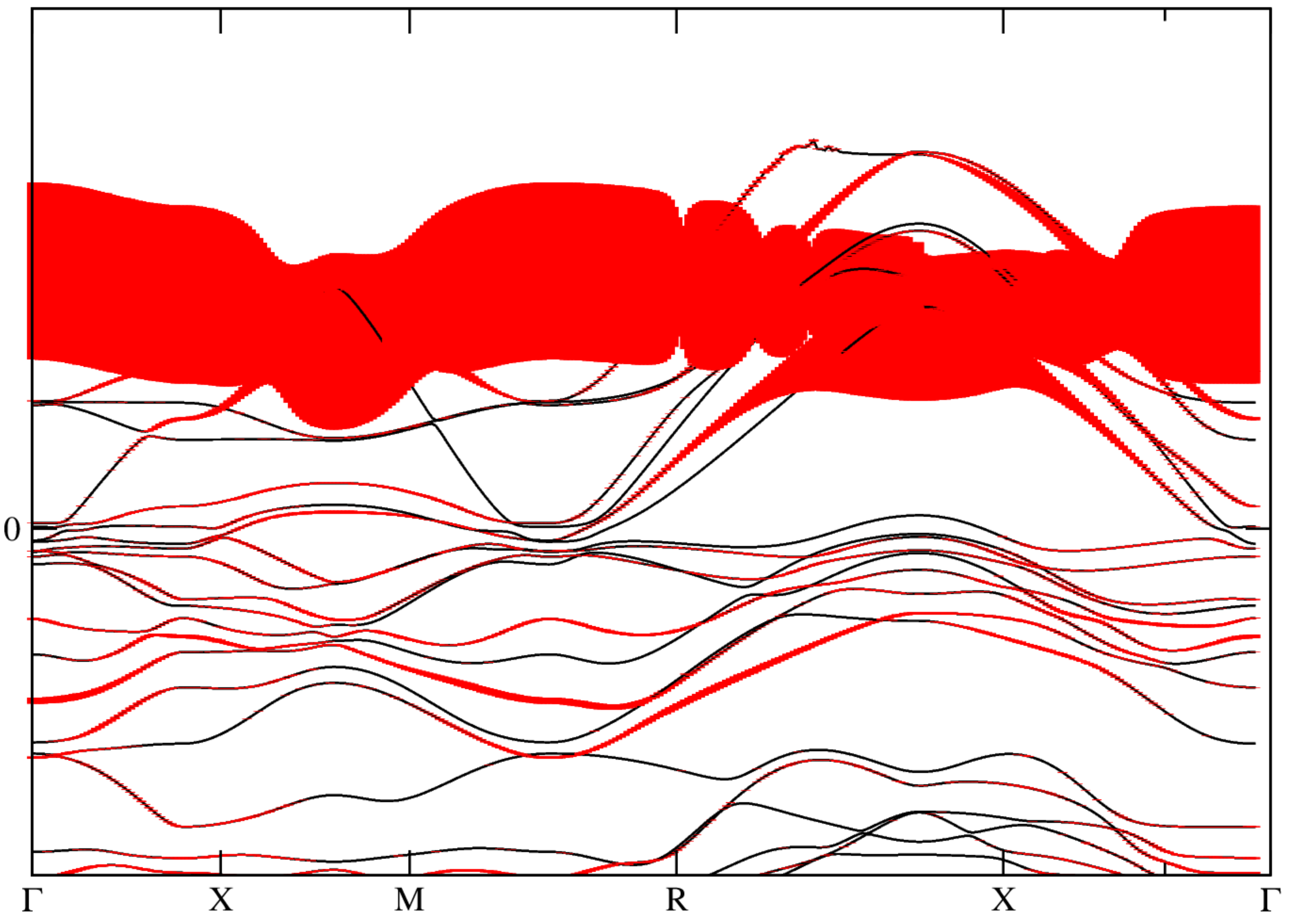}}
\subfigure[]{\includegraphics [width=6cm]{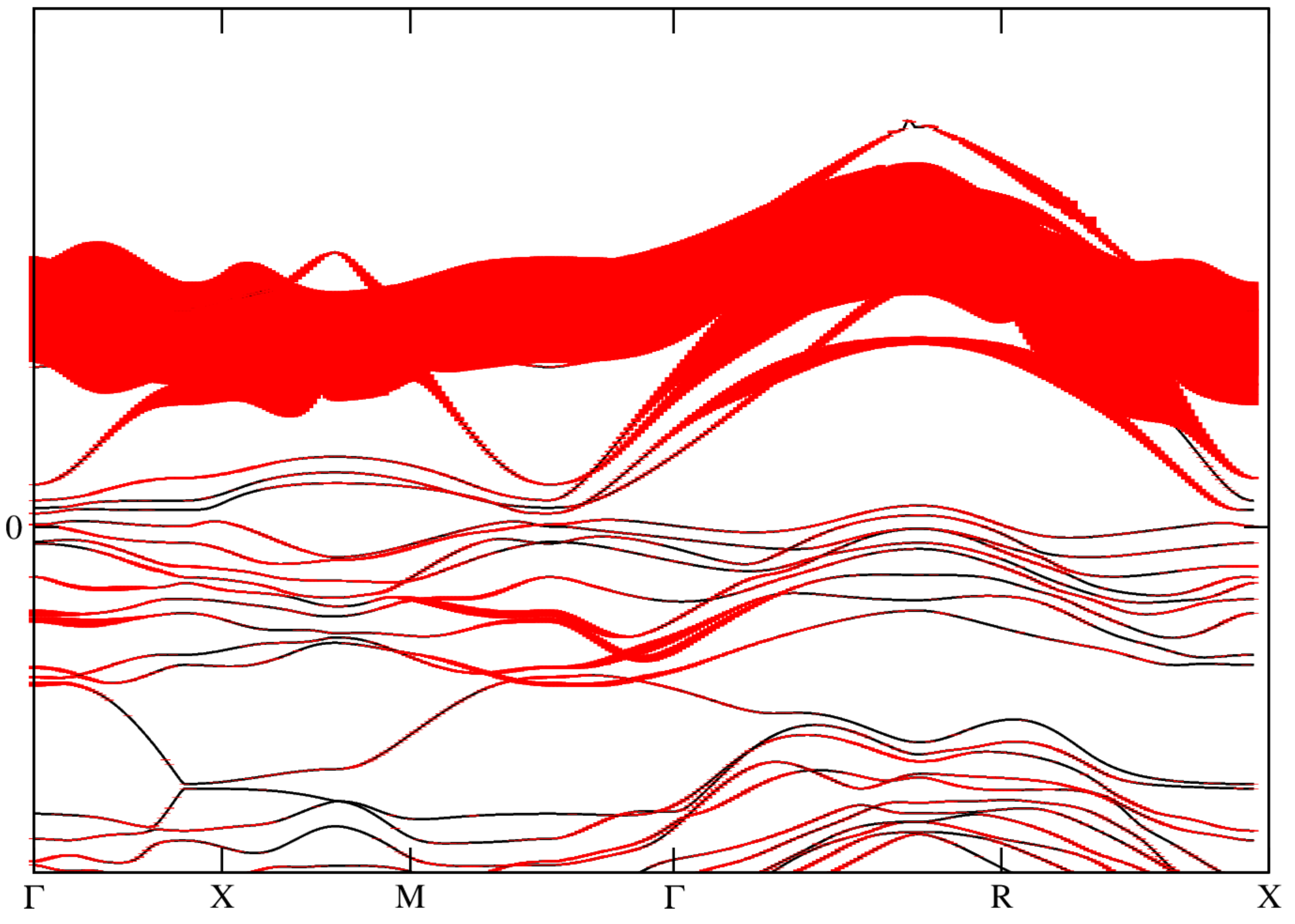}}
\subfigure[]{\includegraphics [width=6cm]{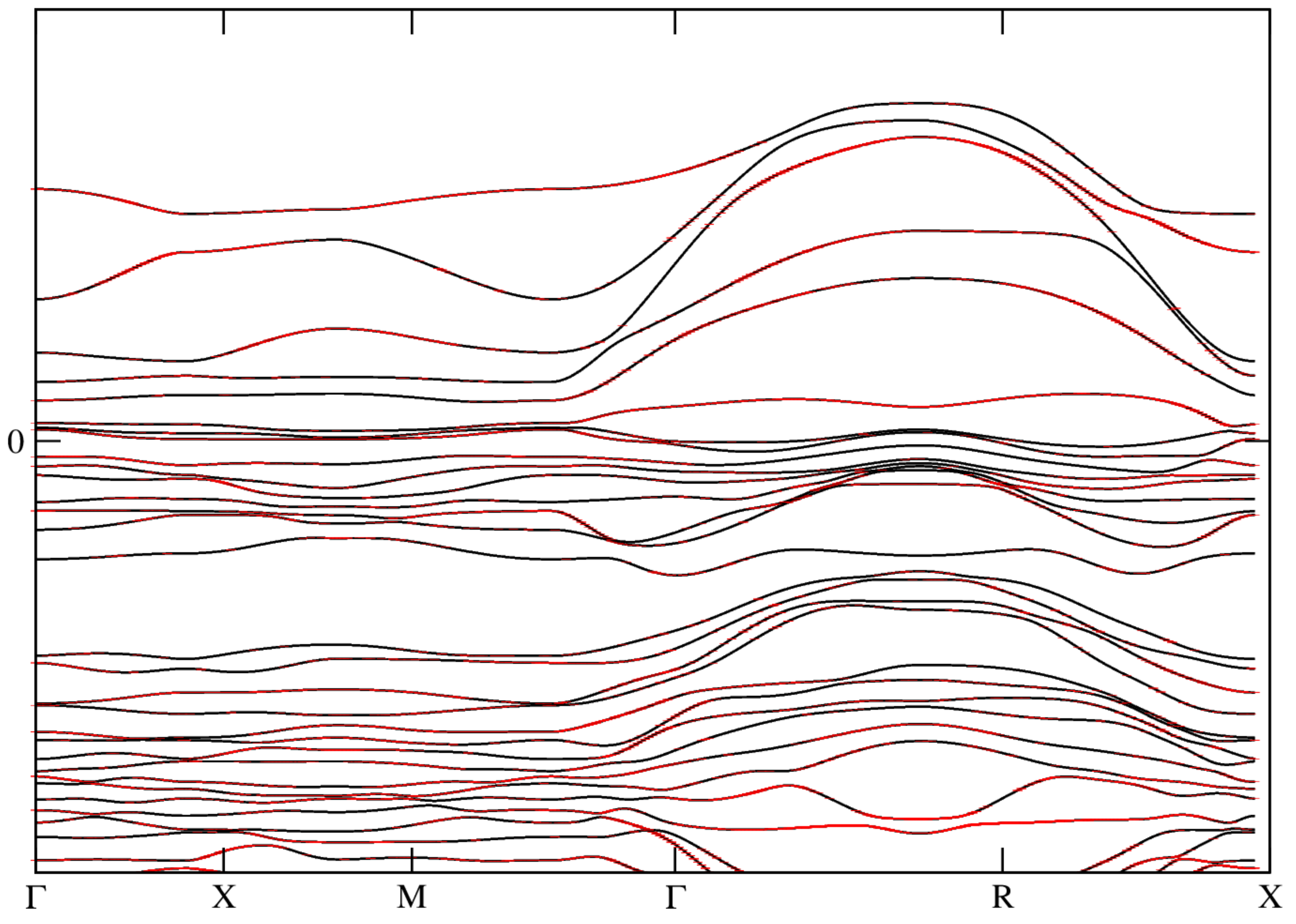}}
\subfigure[]{\includegraphics [width=6cm]{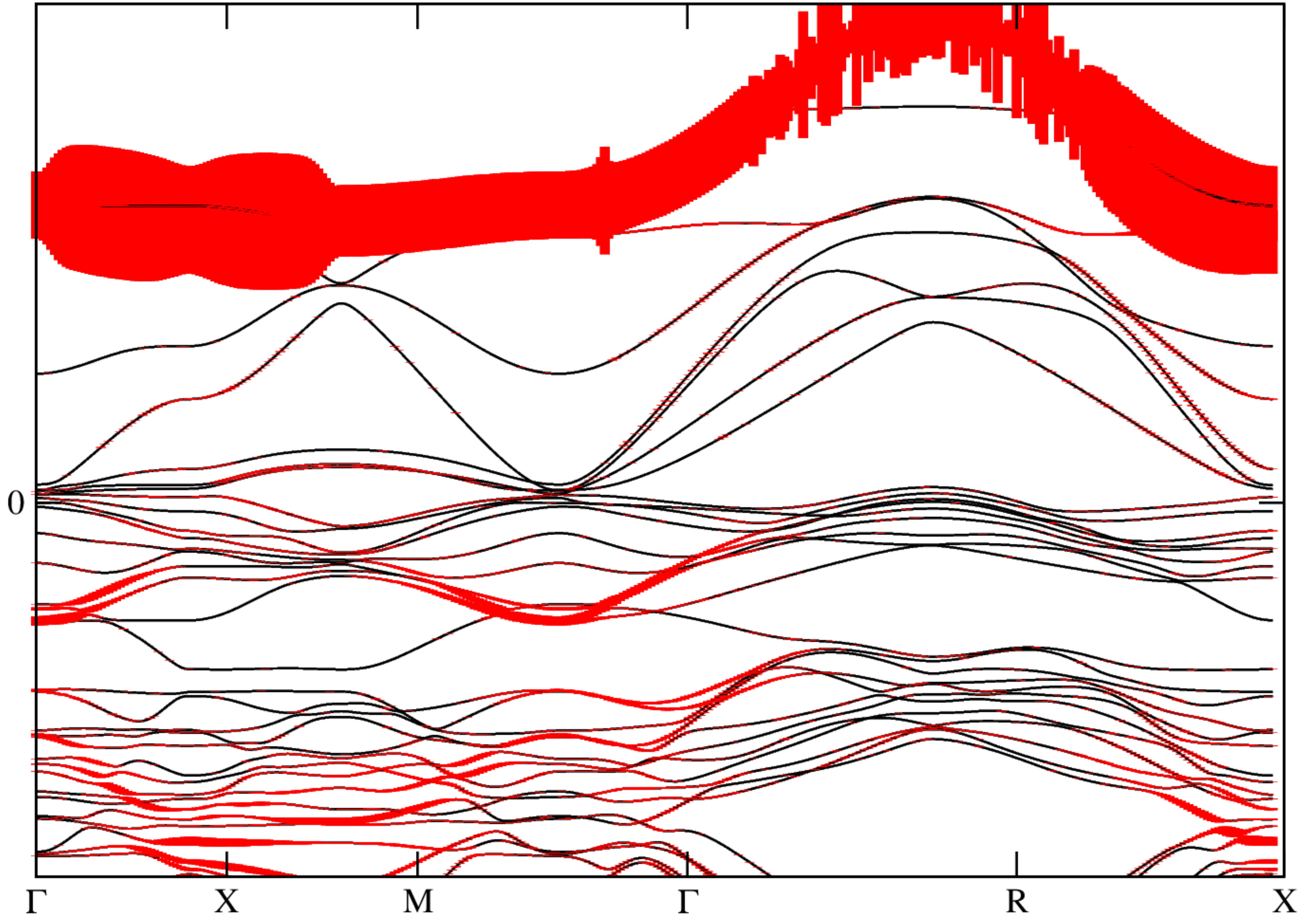}}
\caption{Fat-band structure for (a) Mo-doped SrCoO$_3$, (b) V-doped SrCoO$_3$, (c) P-doped SrCoO$_3$, (d) Nb-doped SrCoO$_3$. The red areas indicate the contribution of dopants to the band structure. The high symmetry k-points for simple cubic structures were chosen to most effectively sample the First Brillouin Zone and are $\Gamma$ (0, 0, 0), X(0, 0.5, 0), M(0.5, 0.5, 0.0), and R(0.5, 0.5, 0.5)}
\label{dopant_band_structure}
\end{figure}

The band structure of all four dopants reveals that there is little to no band gap around the Fermi level as expected, which reveals the metallic behavior of these systems. The contributions of the dopant to the band structure is shown in Figure \ref{dopant_band_structure} in the form of fat-bands in order to analyze how the dopant affects the band structure. Figure \ref{dopant_band_structure} shows that niobium, vanadium and molybdenum substantially affect the band structure of the complex, while Phosphorus does so to a lesser extent because of the lack of a $d$ orbital. This reveals that metallic dopants are more effective than non-metallic dopants for supercapacitor applications. Therefore, we can conclude that metallic dopants are more effective than nonmetallic dopants for supercapacitor electrodes.\par

\begin{figure}
\includegraphics [width=10cm]{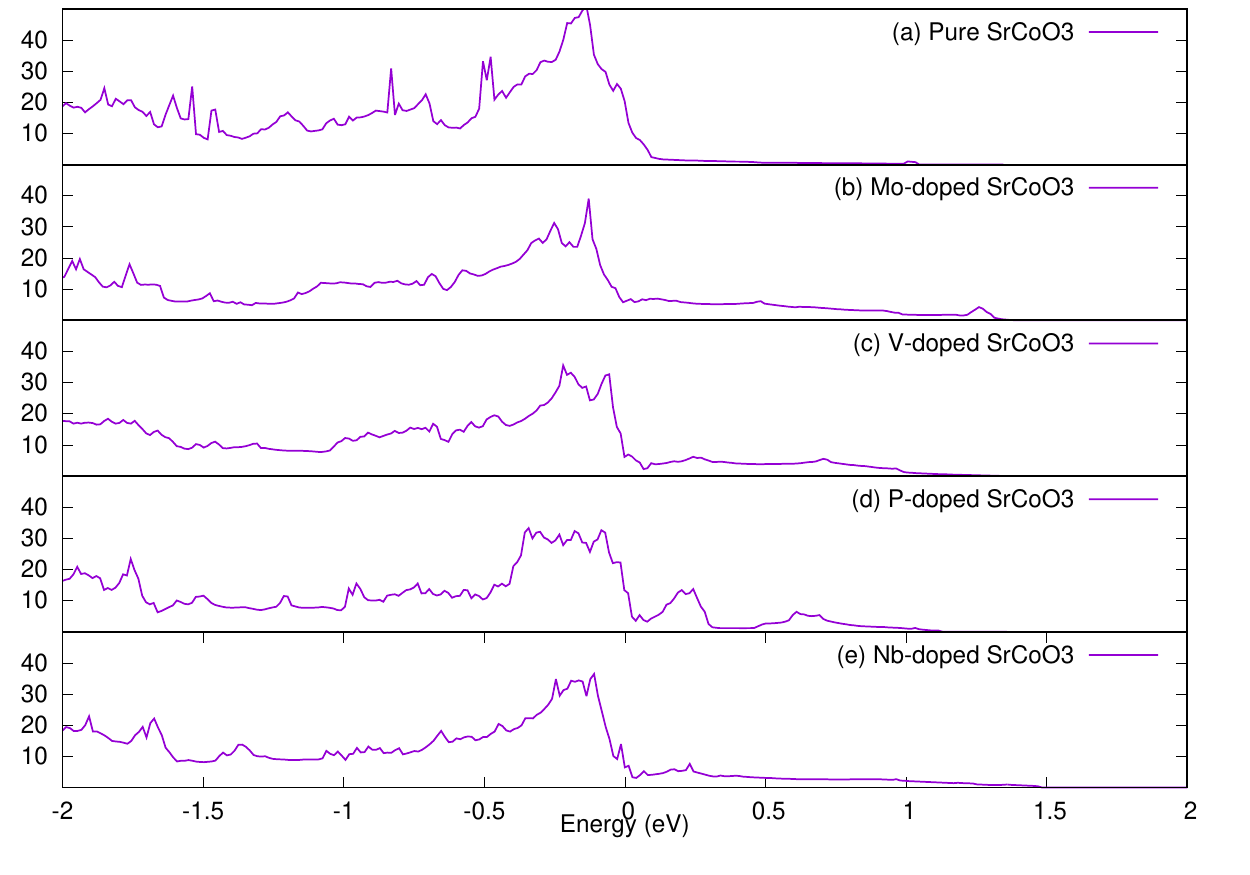}
\caption{TDOS for all of the doped compounds. Figure (a) shows the TDOS of pure SrCoO$_3$, (b) shows the TDOS for Mo-doped SrCoO$_3$, (c) shows the TDOS for V-doped SrCoO$_3$, (d) shows the TDOS for P-doped SrCoO$_3$, and (e) shows the TDOS for Nb-doped SrCoO$_3$.}
\label{dopant_tdos}
\end{figure}

The total DOS was analyzed for further insight into the conductive behavior of the dopants and how they affect the total conductivity of the complex. The TDOS was calculated by first calculating the total energy using the converged kinetic energy cutoffs and k-point grids, along with the relaxed lattice constants of the complexes, and is plotted in Figure \ref{dopant_tdos}. The total DOS of pure SrCoO$_3$ is also plotted in Figure \ref{dopant_tdos} (e) for comparison.

\begin{figure}
\subfigure[]{\includegraphics [width=7cm]{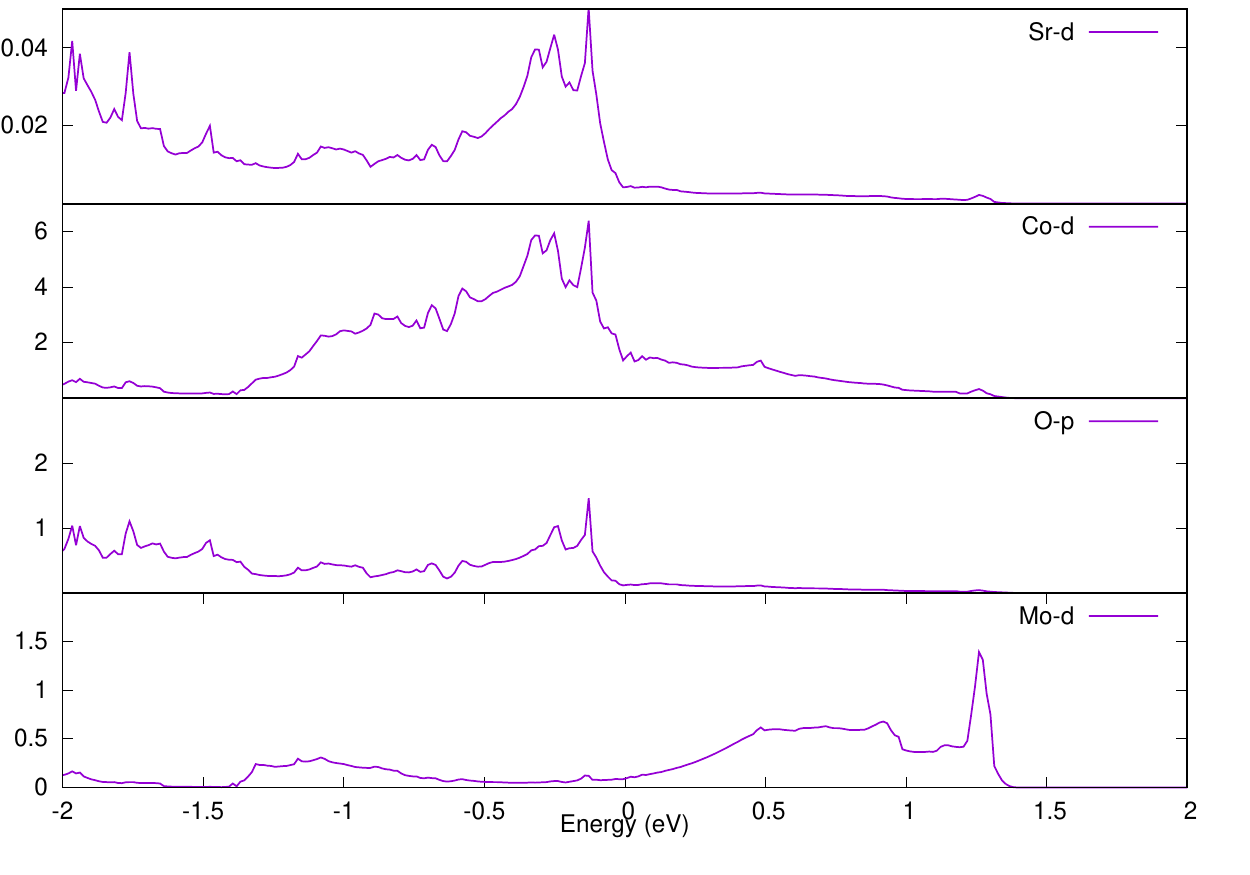}}
\subfigure[]{\includegraphics [width=7cm]{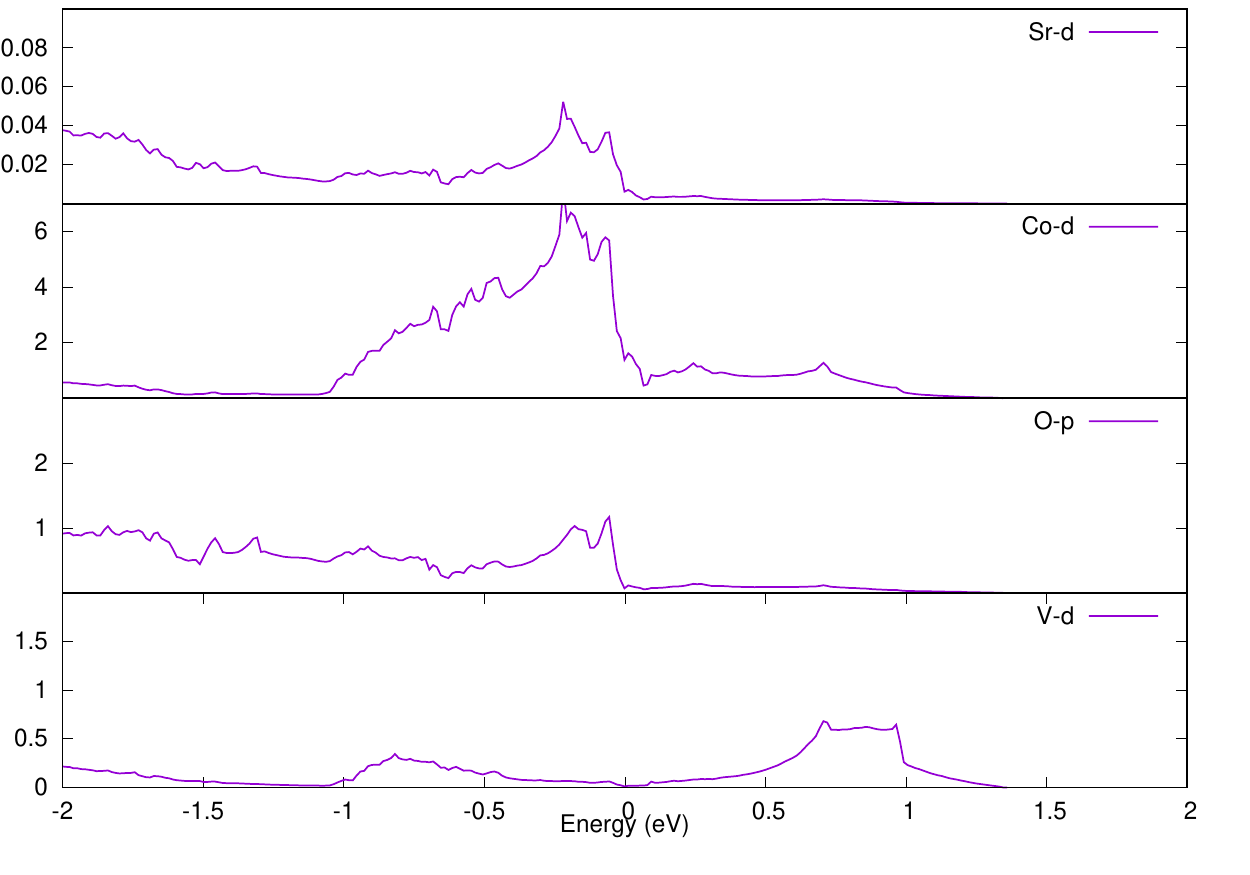}}
\subfigure[]{\includegraphics [width=7cm]{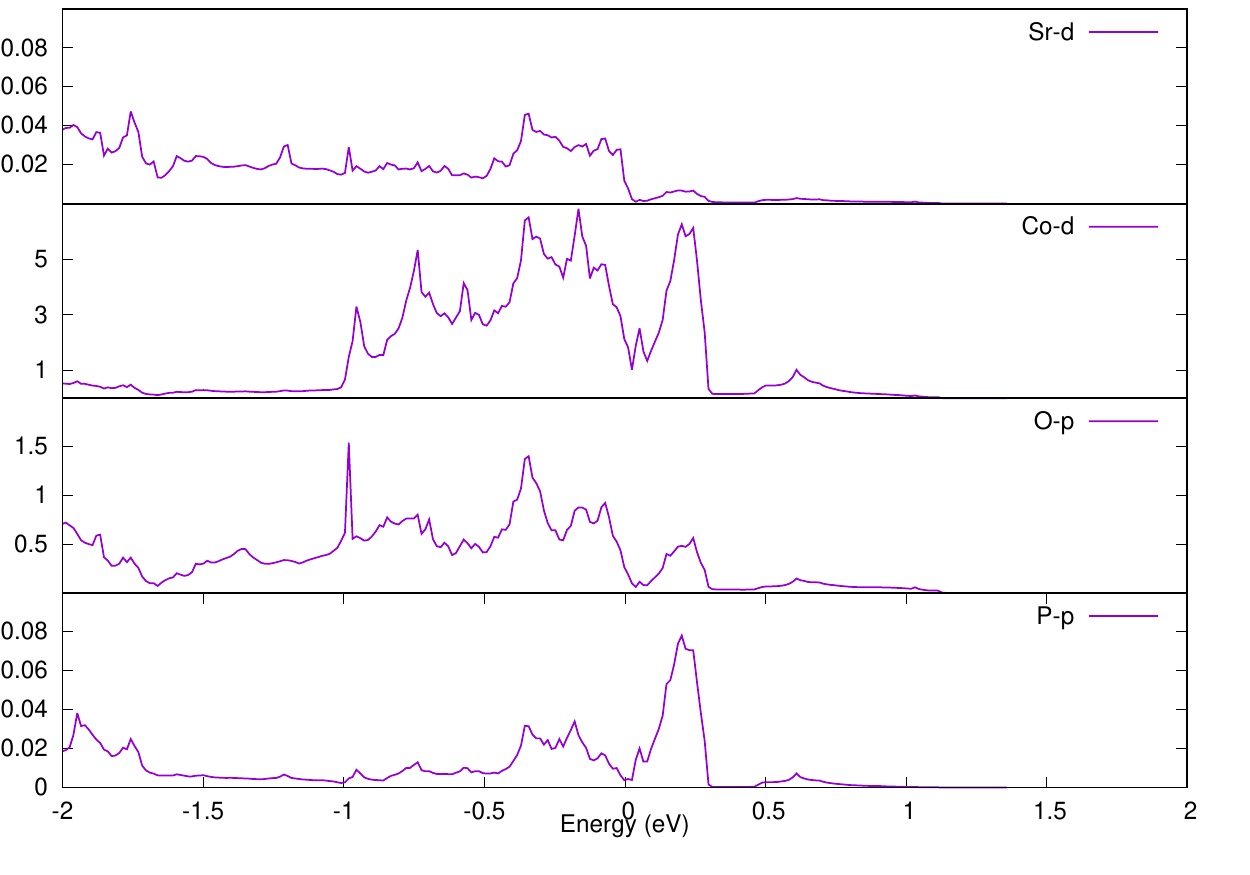}}
\subfigure[]{\includegraphics [width=7cm]{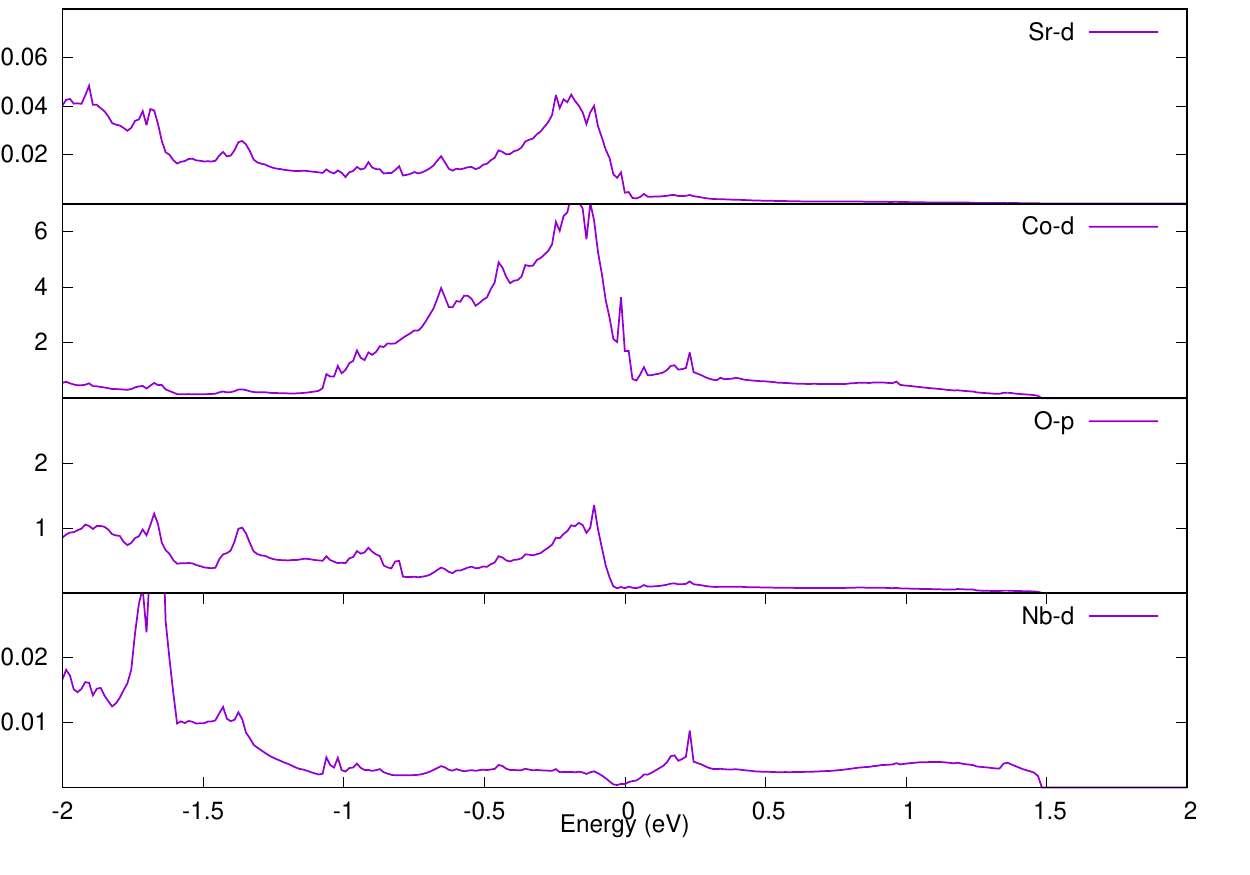}}
\caption{PDOS for all of the doped compounds. For O, the 2 $p$ orbital was analyzed, and the 3 $d$ orbital was plotted for Co and Sr. For Molybdenum, shown in (a), the 4 $d$ orbital was plotted, and for Vanadium, shown in (b), the 3 $d$ orbital was plotted. Phosphorus does not have a $d$ orbital, so the 2 $p$ orbital was analyzed instead in (c). The 4 $d$ orbital was plotted for Niobium in (d).}
\label{dopant_pdos}
\end{figure}

Analysis of the TDOS and PDOS reveals the conductive behavior of doped SrCo$O_3$. Figure \ref{dopant_pdos} reveals that all four dopants increase the conductivity substantively, as shown by the non-zero localized states around the Fermi level. However, it is clear that phosphorus is the least conductive dopant because the values of the states are considerably lower for phosphorus TDOS than for any of the other dopants. The PDOS shows that Mo and V achieve similar results as dopants, with an increased conductivity shown by the $3p$-orbital for Sr, the $d$-orbital for Cobalt, the $2p$-orbital for Oxygen, and the $d$-orbital for the dopants. It can be said that the increased number of peaks shown in the PDOS are a result of the hybridization of the Sr $3d$, Co $3d$, O $2p$ and the $3d$ orbitals of the metallic dopants. It is evident that Niobium is by far the best dopant for SrCoO$_3$, as the states are much higher and there are more states for Nb-doped SrCoO$_3$ than any other dopant. For all four dopants, it is evident from the nonzero states around the fermi level that there is an overlap between the conduction bands (CB) and valence bands (VB) and that the states shift away from the CB towards the VB. For all of the dopants, there are states around the Fermi level for all four elements (Sr, Co, O, and the dopant), showing that all of the elements contribute equally to the conduction process.  Taken together, the formation energy, band structure and PDOS reveal that while all of the dopants studied increase the compound's conductivity, niobium is the best dopant due to its higher stability (as given by the lower formation energy) and its greater conductivity (as shown by the PDOS and the band structure). Furthermore, the PDOS and fat-band structure reveal that phosphorus is not as effective as the rest of the dopants. This shows that metallic dopants should be used to improve supercapacitor performance. In addition, we have discovered that vanadium is a promising new dopant for SrCoO$_3$ as anion-intercalation-type supercapacitor electrode. \par
Our findings suggest the viability of a novel SrCoO$_3$$/$graphene interface for supercapacitor applications. Previous studies involving TMO/graphene interfacial structures have suggested that oxygen acts as a conducting channel between graphene and the cluster structure\cite{hussain2019, xiong2013}, which increases the conductivity of the interface, which supports our electronic charge density calculations. Compared to FeO$/$ graphene, which has a formation energy of 3.7 eV\cite{hussain2019}, and MnO$_2$$/$graphene, which has a formation energy of 5.5 eV\cite{xiong2013}, we find that our SrCoO$_3$$/$graphene interface is relatively stable with a formation energy of only 1.3 eV. Our PDOS calculations are also supported by Xiong et al\cite{xiong2013} and Hussain et al\cite{hussain2019}, who report that TMO/graphene interfacial structures exhibit more peaks that coincide with each other due to the strong hybridization of the Sr $3d$, Co $3d$, C $2p$ and O $2p$ orbitals. \par
We also theoretically verify that inducing oxygen vacancies in SrCoO$_3$ increases the conductivity of SrCoO$_3$ due to the strong hybridization of the Sr $3d$, Co $3d$, and O $2p$ orbitals. Yang et al \cite{yang2018} and Cheng et al \cite{cheng2017}, among others, have concluded that inducing oxygen vacancies in TMOs decreases the stability but increases the conductivity of the material, which we confirm through formation energy and PDOS calculations. In addition, previous studies have also found that inducing oxygen vacancies results in charge accumulation around the vacancy due to electron delocalization in the surrounding atoms\cite{cheng2017, cheng2013} this is supported by our electronic charge density calculations. \par
Furthermore, we find that doping at the B-site with Mo, V, P, and Nb increases the stability of this material and also improves the conductivity of SrCoO$_3$. Mo has been previously studied as a dopant for SrCoO$_3$ in both supercapacitor applications\cite{tomar2018} and for solid oxide fuel cells\cite{aguadero2012} due to its OER-enhancing properties, which make this perovskite more conductive. We confirm this conductive behavior through fat-band calculations, which reveal Molybdenum's increased contribution to the band structure of Mo-doped SrCoO$_3$. Our PDOS calculations show that Mo-doped SrCoO$_3$ results in strong hybridization of Mo $d$, Sr $d$, Co $d$, and O $p$ orbitals. Zhu et al\cite{zhu2016} studied the effect of P-doping on SrCoO$_3$ as an OER catalyst and concluded that phosphorus significantly changes the structure of SrCoO$_3$, which supports our relaxation calculations. Although they concluded that this resulted in a more stable compound, we find that the other dopants increase the stability of SrCoO$_3$ to a much greater extent. However, previous studies have concluded that a small amount of doping at the B-site increases the stability of SrCoO$_3$\cite{george2018, nan2019}. We confirm this as doping with any of the four dopants resulted in a complex with a very low formation energy, indicating the stability. Li et al\cite{li2017niobium} found that doping SrCoO$_3$ with niobium resulted in a very high energy density and also prolonged cycling life, which supports our conclusion that Niobium is the best dopant of the four dopants because it significantly increases the stability of this compound and also contributes tremendously to the conductivity. To the best of our knowledge, vanadium and phosphorus have not been studied as dopants for SrCoO$_3$ for supercapacitor applications. Nevertheless, our conclusions that these compounds enhance the OER of SrCoO$_3$ and also increase the stability are supported by the findings of Zhu et al\cite{zhu2016} and Guo et al\cite{guo2018}. However, we find that phosphorus is the least effective dopant due to its lack of a $d$ orbital, which reveals that metallic dopants are more effective for anion-intercalation-type supercapacitors. Therefore, we can conclude that vanadium and to a lesser degree phosphorus are both novel, promising dopants for supercapacitor applications.  
\newpage
\section{Conclusion}
In this study, we use Density Functional Theory to explore various methods the performance of the perovskite SrCoO$_3$ for supercapacitor applications. We systematically study electronic structure properties and formation energies. \par
An SrCoO$_3$$/$graphene interface was studied to analyze how graphene contributes to the performance of SrCoO$_3$ as a supercapacitor electrode. The formation energy of the interface was 1.3 eV, which is relatively stable compared to other TMO$/$graphene interface structures. Both DOS calculations and electronic charge density indicate that the interface is extremely conductive. \par
Oxygen vacancies were induced to determine the mechanism by which vacancies improve the conductivity of SrCoO$_3$. The defect formation energy of one oxygen vacancy is 2.5 eV, which is comparable to other studies which have induced oxygen vacancies in TMOs. Fat-band structures reveal that vacancies cause the oxygen $2p$ orbital to contribute more to the band structure, and DOS calculations support this conclusion. The PDOS calculations reveal that oxygen vacancies cause a dramatic increase in the conductivity of the material. \par
We also doped SrCoO$_3$ with four different dopants, Mo, V, P, and Nb, at 25\% concentrations at the B-site by replacing one Co atom with one dopant atom. We evaluated the performance of each dopant by calculating the formation energy, TDOS, PDOS, and fat-band structure. All calculations show that all four dopants improve SrCoO$_3$ performance drastically. Niobium is the best dopant of the four because it had the lowest formation energy of -9.89 eV. In addition, Niobium contributed the most to the fat-band structure of any of the four dopants and is the most conductive as shown by PDOS. Vanadium is also a promising new dopant for anion-intercalation-type supercapacitor electrodes. \par
Therefore, we conclude that a graphene interface, oxygen vacancies, and doping can all improve the performance of SrCoO$_3$ as an anion-intercalation type supercapacitor electrode to different extents.
\newpage
\bibliography{my_ref}{}
\bibliographystyle{apsrev4-1}

\end{document}